\documentclass[apj,numberedappendix]{emulateapj}

\slugcomment{To Appear in {\em Publications of the Astronomical Society of the Pacific}}

\shorttitle{CalFUSE: The \fuse\/ Calibration Pipeline}
\shortauthors{Dixon et al.}

\newcommand{\eg}{e.g.}

\newcommand{\etc}{etc.}
\newcommand{\fig}[1]{Fig.~\ref{#1}}

\newcommand{\ie}{i.e.}

\newcommand{\cthree}{\ion{C}{3}}

\newcommand{\flux}{erg cm$^{-2}$ s$^{-1}$  \AA$^{-1}$}

\newcommand{\htwo}{H$_2$}
\newcommand{\kms}{km s$^{-1}$}

\newcommand{\ntwo}{\ion{N}{2}}
\newcommand{\nthree}{\ion{N}{3}}

\newcommand{\oone}{\ion{O}{1}}
\newcommand{\osix}{\ion{O}{6}}

\newcommand{\fuse}{{\it FUSE}}
\newcommand{\hst}{{\it HST}}

\begin{document}
 
\title{CalFUSE \lowercase{v3}: A Data-Reduction Pipeline for the {\em Far Ultraviolet Spectroscopic
Explorer}\footnotemark[1]}

\footnotetext[1]{Based on observations made with the NASA-CNES-CSA {\it Far Ultraviolet Spectroscopic Explorer. FUSE}\/ is operated for NASA by the Johns Hopkins University under NASA contract NAS5-32985.}

\author{W.~V. Dixon\altaffilmark{2},
D.~J. Sahnow\altaffilmark{2},
P.~E. Barrett\altaffilmark{3,4},
T. Civeit\altaffilmark{2,5},
J. Dupuis\altaffilmark{2,6}, 
A.~W. Fullerton\altaffilmark{2,7},
B.~Godard\altaffilmark{2,8},
J.-C.~Hsu\altaffilmark{3,9},
M.~E. Kaiser\altaffilmark{2},
J.~W. Kruk\altaffilmark{2},
S. Lacour\altaffilmark{2,10},
D.~J. Lindler\altaffilmark{11},
D.~Massa\altaffilmark{12},
R.~D. Robinson\altaffilmark{2,9},
M.~L.~Romelfanger\altaffilmark{2}, and
P. Sonnentrucker\altaffilmark{2}
}

\altaffiltext{2}{Department of Physics and Astronomy,
Johns Hopkins University, 3400 N. Charles Street, Baltimore, MD 21218}

\altaffiltext{3}{Space Telescope Science Institute, ESS/SSG, 
3700 San Martin Drive, Baltimore, MD 21218}

\altaffiltext{4}{Current address: Earth Orientation Department,
U.S. Naval Observatory,
3450 Massachusetts Avenue NW,
Washington, DC 20392}

\altaffiltext{5}{Primary affiliation: Centre National d'\'{E}tudes Spatiales, 2 place Maurice Quentin, 75039 Paris Cedex 1, France}

\altaffiltext{6}{Current address: Canadian Space Agency,
6767 route de l'A\'{e}roport, Longueuil, QC, Canada, J3Y 8Y9}

\altaffiltext{7}{Primary affiliation: Department of Physics and Astronomy, University of Victoria,
P.~O. Box 3055, Victoria, BC V8W~3P6, Canada}

\altaffiltext{8}{Current address: Institut d'Astrophysique de Paris, 98 bis, boulevard Arago, 75014 Paris, France}

\altaffiltext{9}{Retired}

\altaffiltext{10}{Current address: Sydney University, NSW 2006, Australia}

\altaffiltext{11}{Sigma Space Corporation, 4801 Forbes Boulevard, Lanham, MD  20706}

\altaffiltext{12}{SGT, Inc., NASA Goddard Space Flight Center, Code 665.0, Greenbelt, MD 20771}

\setcounter{footnote}{12}

\begin{abstract}

Since its launch in 1999, the {\it Far Ultraviolet Spectroscopic
Explorer (FUSE)}\/ has made over 4600 observations of some 2500 individual targets.
The data are reduced by the Principal Investigator team at the Johns Hopkins
University and archived at the Multimission Archive at Space
Telescope (MAST).  The data-reduction software package, called CalFUSE,
has evolved considerably over the lifetime of the mission.  
The entire \fuse\/ data set has recently been reprocessed
with CalFUSE v3.2, the latest version of this software.  This paper 
describes CalFUSE v3.2, the instrument calibrations upon which
it is based, and the format of the resulting calibrated data files.

\end{abstract}
 

\section{INTRODUCTION}

The {\it Far Ultraviolet Spectroscopic Explorer (FUSE)}\/ is a high-resolution, far-ultraviolet spectrometer operating in the 905--1187 \AA\ wavelength range.  \fuse\/ was launched in 1999 on a Delta II rocket into a nearly circular, low-earth orbit with an inclination of 25\degr\ to the equator and an approximately 100-minute orbital period.
Data obtained with the instrument are reduced by the principal investigator team at the Johns Hopkins University using a suite of computer programs called CalFUSE.  Both raw and processed data files are deposited in the \anchor{http://archive.stsci.edu/fuse}{Multimission Archive at Space Telescope (MAST)}.

CalFUSE evolved considerably in the years following launch as our increasing knowledge of the spectrograph's performance allowed us to correct the data for more and more instrumental effects.  The program eventually became unwieldy, and in 2002 we began a project to re-write the code, incorporating our new understanding of the instrument and best practices for data reduction.
The result is CalFUSE v3, which produces a higher quality of calibrated data while running ten times faster than previous versions.  The entire \fuse\/ archive has recently been reprocessed with CalFUSE v3.2; we expect this to be the final calibration of these data.

In this paper, we describe CalFUSE v3.2.0 and its calibrated data products.
Because this document is meant to serve as a resource for researchers
analyzing archival \fuse\/ spectra,
we emphasize the interpretation of processed data files obtained from MAST rather than the details of designing or running the pipeline. 
An overview of the \fuse\/ instrument is provided in \S\ \ref{sec_fuse}, 
and an overview of the pipeline in \S\ \ref{sec_calfuse}.
Section \ref{sec_stepstep} presents a detailed description of the pipeline modules and their subroutines.  
The \fuse\/ wavelength and flux calibration are discussed in \S\ \ref{sec_calfiles},
and a few additional topics are considered in \S\ \ref{sec_discussion}.
A detailed description of the various file formats employed by CalFUSE is presented in the Appendix.

Additional documentation available from MAST includes 
the CalFUSE Homepage,\footnote{\url{http://archive.stsci.edu/fuse/calfuse.html}}
{{\em The CalFUSE Pipeline Reference Guide}},\footnote{\url{http://archive.stsci.edu/fuse/pipeline.html}}
{{\em The FUSE Instrument and Data Handbook}},\footnote{\url{http://archive.stsci.edu/fuse/dhbook.html}}
and 
{{\em The FUSE Data Analysis Cookbook}}.\footnote{\url{http://archive.stsci.edu/fuse/cookbook.html}}

\section{THE {\em FUSE}\/ INSTRUMENT}\label{sec_fuse}

\begin{figure}
\epsscale{.60}
\plotone{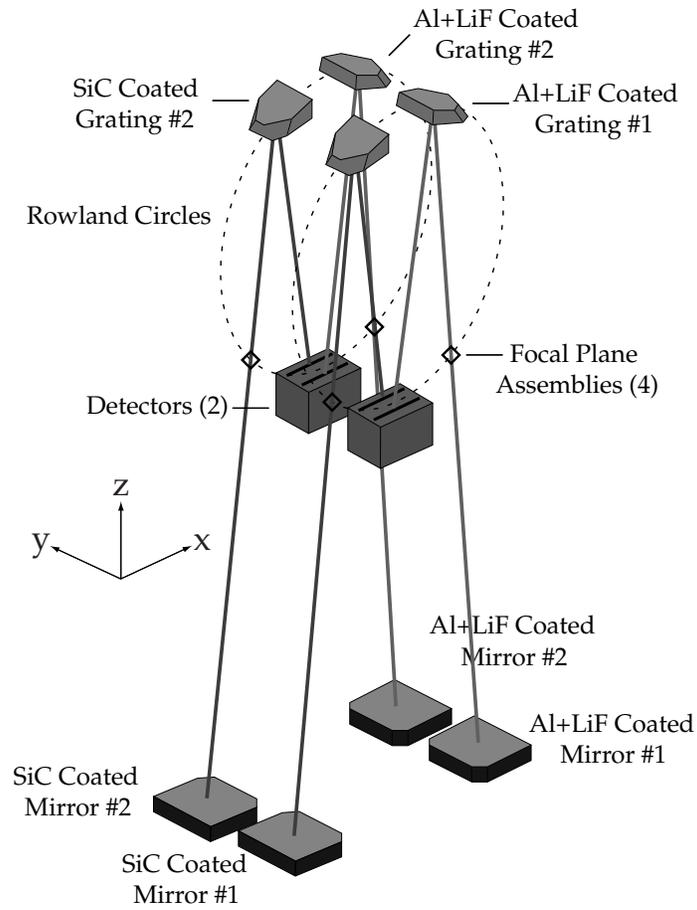}
\caption{Schematic of the \fuse\/ instrument optical system.  The telescope focal lengths are 2245 mm, and the Rowland circle diameters are 1652 mm.  (Figure from Moos et al.\ 2000.)}
\label{toaster}
\end{figure}

\fuse\/ consists of four co-aligned prime-focus telescopes, each with its own
Rowland spectrograph (\fig{toaster}).
Two of the four channels employ Al+LiF optical coatings and record spectra over the wavelength range $\sim$ 990--1187 \AA, while the other two use SiC coatings, which provide reflectivity to wavelengths below the Lyman limit.
The four channels overlap between 990 and 1070~\AA.
Spectral resolution is roughly 20,000 ($\lambda / \Delta \lambda$) for point sources.
For a complete description of \fuse, see \citet{Moos:00} and \citet{Sahnow:00}.

\begin{figure}
\epsscale{.70}
\plotone{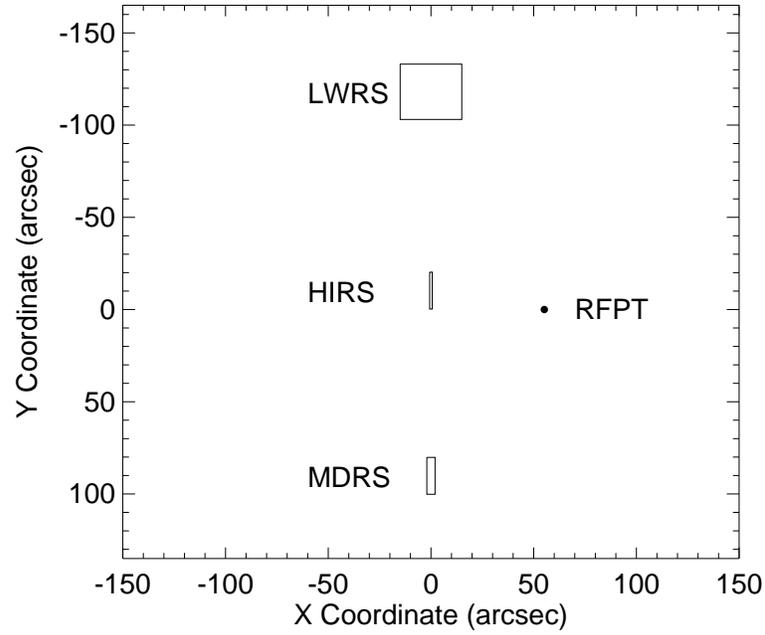}
\caption{The \fuse\/ apertures projected onto the sky.  In the FPA coordinate system, the LWRS, HIRS, and MDRS apertures are centered at Y = $-$118\farcs07, $-10$\farcs27, and +90\farcs18, respectively.  The reference point (RFPT) at X = +55\farcs18 is not an aperture; when a target is placed at this location, the three apertures sample the background sky.  With north on top and east on the left, this diagram corresponds to an aperture position angle of 0\arcdeg.  Positive aperture position angles correspond to a counter-clockwise rotation of the spacecraft about the target aperture.  This diagram represents only a portion of the FPA; its active area is 19\arcmin$ \times $19\arcmin.}
\label{fpa}
\end{figure}

At the prime focus of each mirror lies a focal-plane assembly (or FPA, shown in \fig{fpa}) containing three spectrograph entrance apertures: the low-resolution aperture (LWRS; 30\arcsec\ $\times$ 30\arcsec), used for most observations, the medium-resolution aperture (MDRS; 4\arcsec\ $\times$ 20\arcsec), and the high-resolution aperture (HIRS; 1.25\arcsec\ $\times$ 20\arcsec).
The reference point (RFPT) is not an aperture; when a target is placed at this location, the three apertures sample the background sky.  For a particular exposure, the FITS file header keywords RA\_TARG and DEC\_TARG contain the J2000 coordinates of the aperture (or RFPT) listed in the APERTURE keyword, while the keyword APER\_PA contains the position angle of the $-$Y axis
(in the FPA coordinate system; see \fig{fpa}), corresponding to a counter-clockwise rotation of the spacecraft about the target (and thus about the center of the target aperture).  

The spectra from the four instrument channels are imaged onto two photon-counting microchannel-plate (MCP) detectors, labeled 1 and 2, with a LiF spectrum and a SiC spectrum on each (\fig{toaster}).  Each detector is comprised of two MCP segments, labeled A and B.  Raw science data from each detector segment are stored in a separate data file; an exposure thus yields four raw data files, labeled 1A, 1B, 2A, and 2B.  Because the three apertures are open to the sky at all times, the LiF and SiC channels each generate three spectra, one from each aperture.  In most cases, the non-target apertures are empty and sample the background sky.  Figure \ref{detector1a} presents a fully-corrected image of detector 1A obtained during a bright-earth observation.  The emission features in all three apertures are geocoronal.  Note that the LiF1 wavelength scale increases to the right, while the SiC1 scale increases to the left.  The Lyman $\beta$ $\lambda 1026$ airglow feature is prominent in each aperture.

Two observing modes are available:
In photon-address mode, also known as time-tag or TTAG mode, the X and Y coordinates and pulse height (\S\ \ref{sec_walk}) of each detected photon are stored in a photon-event list.  A time stamp is inserted into the data stream, typically once per second.  Data from the entire active area of the detector are recorded. 
Observing bright targets in time-tag mode can rapidly fill the spacecraft recorder. Consequently, when a target is expected to generate more than $\sim$ 2500 counts s$^{-1}$ across all four detector segments, the data are stored in spectral-image mode, also called histogram or HIST mode. To conserve memory, histogram data are (usually) binned by eight pixels in Y (the spatial dimension), but unbinned in X (the dispersion dimension).  Only data obtained through the target aperture are recorded.  Individual photon arrival time and pulse height information is lost.  The orbital velocity of the \fuse\/ spacecraft is 7.5 \kms. Since Doppler compensation is not performed by the detector electronics, histogram exposures must be kept short to preserve spectral resolution; a typical histogram exposure is about 500 s in length.

The front surfaces of the FPAs are reflective in visible light.  On the two LiF channels, light not passing through the apertures is reflected into a visible-light CCD camera. Images of stars in the field of view around the apertures are used for acquisition and guiding by this camera system, called the Fine Error Sensor (FES).  \fuse\/ carries two redundant FES cameras, which were provided by the Canadian Space Agency.  
FES A views the FPA on the LiF1 channel, and FES B views the LiF2 FPA.  
During initial checkout, FES A was designated the default camera and was used for all science observations until it began to malfunction in 2005.  
In July of that year, FES B was made the default guide camera.
Implications of the switch from FES A to FES B are discussed in \S\ \ref{sec_fesb}.

\section{OVERVIEW OF CALFUSE}\label{sec_calfuse}

The new CalFUSE pipeline was designed with three principles mind: the first was that, to the extent possible, we follow the path of a photon backwards through the instrument, correcting for the instrumental effects introduced in each step.  The principal steps in this path, together with the effects imparted by each, are listed below.  Most of the optical and electronic components in this list are labeled in \fig{toaster}.  

\noindent
1.	Satellite motion imparts a Doppler shift.\\
2.	Satellite pointing instabilities shift the target image within (or out of) the aperture.\\
3.	Thermally-induced mirror motions shift the target image within (or out of) the aperture.\\
4.	FPA offsets shift the spectrum on the detector.\\
5.	Thermally-induced motions of the spectrograph gratings shift the target image within (or out of) the aperture.\\
6.	Ion-repelling wire grids can cast shadows called ``worms.''\\
7.	Detector effects include quantum efficiency, flat field, dead spots, and background.\\
8.	The spectra are distorted by temperature-, count-rate, time-, and pulse-height-dependent errors in the photons' measured X and Y coordinates, as well as smaller-scale geometric distortions in the detector image.\\
9.	Count-rate limitations in the detector electronics and the IDS data bus are sources of dead time.

To correct for these effects, we begin at the bottom of the list and (to the extent possible) work backwards.  
First, we adjust the photon weights to account for data lost to dead time (9) and correct the photons'  X and Y coordinates for a variety of detector distortions (8).
Second, we identify periods of unreliable, contaminated, or missing data.
Third, we correct the photons' X and Y coordinates for grating (5), FPA (4), mirror (3), and spacecraft (2) motions.
Fourth, we assign a wavelength to each photon based on its corrected X and Y coordinates (5), then convert to a heliocentric wavelength scale (1).
Finally, we correct for detector dead spots (7); model and subtract the detector and scattered-light backgrounds (7); and extract (using optimal extraction, if possible), flux calibrate (7) and write to separate FITS files the target's LiF and SiC spectra.
Note that we cannot correct for the effects of worms (6) or the detector flat field (7).

Our second principal was to make the pipeline as modular as possible.
CalFUSE is written in the C programming language and runs on the Solaris, 
Linux, and Mac OS X (versions 10.2 and higher) operating systems.  
The pipeline consists of a series of modules called by a shell script.
Individual modules may be executed from the command line.
Each performs a set of related corrections (screen data, remove motions, \etc) by calling a series of subroutines.

Our third principal was to maintain the data as a photon list (called an intermediate data file, or IDF) until the final module of the pipeline.  
Input arrays are read from the IDF at the beginning of each module, and output arrays are written at the end.
Bad photons are flagged but not discarded, so the user can examine, filter, and combine processed data files without re-running the pipeline.
Like all \fuse\/ data, IDFs are stored as FITS files \citep{Hanisch:01}; the various
file formats employed by CalFUSE are described in the Appendix.

A \fuse\/ observation consists of a set of exposures obtained with a particular target
in a particular aperture on a particular date.  
Each exposure generates four raw data files,
one per detector segment, and each raw data file yields a 
pair of calibrated spectra (LiF and SiC), for a total of 8 calibrated
spectral files per exposure.  
Each raw data file is processed individually by the pipeline.  
Error and status messages are written to a trailer file (described in \S\ \ref{sec_trailer}).
Spectra are extracted only for the target aperture and are binned in wavelength.
Binning can be set by the user, but the default is 0.013 \AA, which corresponds to about two detector pixels or one fourth of a point-source resolution element.
After processing, additional software is used to generate a set of
observation-level spectral files, the  ALL, ANO, and NVO files described in 
\S\ \ref{obs_files}.  
A complete list of \fuse\/ data files and file-naming conventions may be found in 
\anchor{http://archive.stsci.edu/fuse/dhbook.html#DataFilenames}{{\em The
FUSE Instrument and Data Handbook.}}
All of the exposures that constitute an observation are processed and archived together.

Investigators who wish to re-process their data may retrieve the CalFUSE source code and all associated calibration files from \anchor{http://archive.stsci.edu/fuse/calfuse.html}{the CalFUSE Homepage}.  Instructions for running the pipeline and detailed descriptions of the calibration files are provided in \anchor{http://archive.stsci.edu/fuse/pipeline.html}{{\em The CalFUSE Pipeline Reference Guide}}.  Note that, within the CalFUSE software distribution, all of the calibration files, including the FUSE.TLE file (\S\ \ref{init}), are stored in the directory v3.2/calfiles, while all of the parameter files, including master\_calib\_file.dat and the screening and parameter files (SCRN\_CAL and PARM\_CAL; \S\ \ref{init}), are stored in the directory v3.2/parmfiles.

\section{STEP BY STEP}\label{sec_stepstep}

In this section, we discuss the pipeline subroutine by subroutine.  
Our goal is to describe the algorithms employed by each subroutine
and any shortcomings or caveats of which the user should be aware.  

\subsection{OPUS}\label{sec_OPUS}

The Operations Pipeline Unified System (OPUS) is the data-processing system used by the Space Telescope Science Institute to reduce science data from the {\it Hubble Space Telescope (HST)}.  We use a \fuse -specific version of OPUS to manage our data processing \citep{Rose:98}.  OPUS ingests the data downlinked by the spacecraft and produces the data files that serve as input to the CalFUSE pipeline. OPUS then manages the execution of the pipeline and the files produced by CalFUSE and calls the additional routines that combine spectra from each channel and exposure into a set of observation-level spectral files.  OPUS reads the \fuse\/ Mission Planning Database (which contains target information from the individual observing proposals and instrument configuration and scheduling information from the mission timeline) to populate raw file header keywords and to verify that all of the data expected from an observation were obtained.

OPUS generates six data files for each exposure.  Four are raw data files (identified by the suffix ``fraw.fit''), one for each detector segment.  One is a housekeeping file (``hskpf.fit'') containing time-dependent spacecraft engineering data.  Included in this file are detector voltages, count rates, and spacecraft-pointing information.  The housekeeping file is used to generate a jitter file (``jitrf.fit''), which contains information needed to correct the data for spacecraft motion during an exposure.  Detailed information on the format and contents of each file is provided in the Appendix.
  
\subsection{Generate the Intermediate Data File}\label{init}

The first task of the pipeline is to convert the raw data file into an intermediate data file (IDF), which maintains the data in the form of a photon-event list.  (The format and contents of the IDF are described in \S\ \ref{idf_format}.)  For data obtained in time-tag mode, the module {\tt cf\_ttag\_init} merely copies the arrival time, X and Y detector coordinates, and pulse-height of each photon event from the raw file to the TIME, XRAW, YRAW, and PHA arrays of the IDF.  A fifth array, the photon weight, is initially set to unity.  Photons whose X and Y coordinates place them outside of the active region of the detector are flagged as described in \S\ \ref{sec_active_region}.  Raw histogram data are stored by OPUS as an image; the module {\tt cf\_hist\_init} converts each non-zero pixel of that image into a single entry in the IDF, with X and Y equal to the pixel coordinates (mapped to their location on the unbinned detector), arrival time set to the mid-point of the exposure, and pulse height set to 20 (possible values range from 0 to 31).  The arrival time and pulse height are modified later in the pipeline.  The photon weight is set to the number of accumulated counts on the pixel, \ie, the number of photons detected on that region of the detector.  
 
The IDF has two additional extensions.  The first contains the good-time intervals (GTIs), a series of start and stop times (in seconds from the exposure start time recorded in the file header) computed by OPUS, when the data are thought to be valid.  For time-tag data, this extension is copied directly from the raw data file.  For histogram data, a single GTI is generated with START = 0 and STOP = EXPTIME (the exposure time computed by OPUS).  The final extension is called the timeline table and consists of 16 arrays containing status flags and spacecraft-position, detector high-voltage, and count-rate parameters tabulated once per second throughout the exposure.  Only the day/night and OPUS bits of the time-dependent status flags are populated (\S\ \ref{idf_format}); the others are initialized to zero.  The spacecraft-position parameters are computed as described below.  The detector voltages and the values of various counters are read from the housekeeping data file.

A critical step in the initialization of the IDF is populating the file-header keywords that describe the spacecraft's orbit and control the subsequent actions of the pipeline.  The names of all calibration files to be used by the pipeline are read from the file master\_calib\_file.dat and written to file-header keywords.  (Keywords for each calibration file are included in the discussion that follows.)  Three sets of calibration files are time-dependent:  the effective area is interpolated from the two files with effective dates immediately preceding and following the exposure start time (these file names are stored in the header keywords AEFF1CAL and AEFF2CAL); the scattered-light model is taken from the file with an effective date immediately preceding the exposure start time (keyword BKGD\_CAL); and the orbital elements are read from the FUSE.TLE file, an ASCII file containing NORAD two-line elements for each day of the mission.  These two-line elements are used to populate both the orbital ephemeris keywords in the IDF file header and the various spacecraft-position arrays in the timeline table.  Finally, a series of data-processing keywords is set to either PERFORM or OMIT the subsequent steps of the pipeline.  Once a step is performed, the corresponding keyword is set to COMPLETE.  Some user control of the pipeline is provided by the screening and parameter files (SCRN\_CAL and BKGD\_CAL), which allow one, for example, to select only night-time data or to turn off background subtraction.  An annotated list of file-header keywords, including the calibration files used by the pipeline, is provided in the \anchor{http://archive.stsci.edu/fuse/dhbook.html#FITSHeader}{{\em FUSE Instrument and Data Handbook}}.

{\it Caveats:}  Occasionally, photon arrival times in raw time-tag data files are corrupted.  When this happens, some fraction of the photon events have identical, enormous TIME values, and the good-time intervals contain an entry with START and STOP set to the same large value.  The longest valid exposure spans 55 ks (though most are $\sim$ 2 ks long).  If an entry in the GTI table exceeds this value, the corresponding entry in the timeline table is flagged as bad (using the ``photon arrival time unknown'' flag; \S\ \ref{idf_format}).  Bad TIME values less than 55 ks will not be detected by the pipeline.

Raw histogram files may also be corrupted.  OPUS fills missing pixels in a histogram image with the value 21865.  The pipeline sets the WEIGHT of such pixels to zero and flags them as bad (by setting the photon's ``fill-data bit''; \S\ \ref{idf_format}).  Occasionally, a single bit in a histogram image pixel is flipped, producing (for high-order bits) a ``hot pixel'' in the image.  The pipeline searches for pixels with values greater than 8 times the average of their neighbors, identifies the flipped bit, and resets it.

One or more image extensions may be missing from a raw histogram file (\S \ref{sec_jitter}).  If no extensions are present, the keyword EXP\_STAT in the IDF header is set to $-1$.  Exposures with non-zero values of EXP\_STAT are processed normally by the pipeline, but are not included in the observation-level spectral files ultimately delivered to MAST (\S\ \ref{obs_files}).  Though the file contains no data, the header keyword EXPTIME is not set to zero.

Early versions of the CalFUSE pipeline did not make use of the housekeeping files, but instead employed engineering information downloaded every five minutes in a special ``engineering snapshot'' file.  That information is used by OPUS to populate a variety of header keywords in the raw data file.  If a housekeeping file is not available, CalFUSE v3 uses these keywords to generate the detector high-voltage and count-rate arrays in the timeline table.  Should these header keywords be corrupted, the pipeline issues a warning and attempts to estimate the corrupted values.  In such cases, it is wise to compare the resulting dead-time corrections (\S\ \ref{deadtime}) with those of other, uncorrupted exposures of the same target.

\subsection{Convert to FARF}\label{convert_to_farf}

The pipeline module {\tt cf\_convert\_to\_farf} is designed to remove detector artifacts.  Our goal is to construct the data set that would be obtained with an ideal detector.  The corrections can be grouped into two categories: dead-time effects, which are system limitations that result in the loss of photon events recorded by the detector, and positional inaccuracies, \ie, errors in the raw X and Y pixel coordinates of individual photon events.  The coordinate system defined by these corrections is called the flight alignment reference frame, or FARF.  Corrected coordinates for each photon event are written to the XFARF and YFARF arrays of the IDF.

\subsubsection{Digitizer Keywords}\label{sec_digitizer}

The first subroutine of this module, {\tt cf\_check\_digitizer}, merely compares a set of 16 IDF file header keywords, which record various detector settings, with reference values stored in the calibration file DIGI\_CAL.  Significant differences result in warning messages being written to both the file header and the exposure trailer file.  Such warning messages should be taken seriously, as data obtained when the detectors are not properly configured are likely to be unusable.  Besides issuing a warning, the program sets the EXP\_STAT keyword in the IDF header to $-2$.

\subsubsection{Detector Dead Time}\label{deadtime}

The term ``dead time'' refers specifically to the finite time interval required by the detector electronics to process a photon event.  During this interval, the detector is ``dead'' to incoming photons.  The term is more generally applied to any loss of data that is count-rate dependent.
There are three major contributions to the effective detector dead time on
\fuse.  The first is due to limitations in the detector electronics,  which at
high count rates may not be able to process photon events as fast as they
arrive.  The correction for this effect is computed separately for each segment
from  the count rate measured at the detector anode by the Fast Event Counter
(FEC) and recorded to the engineering data stream, typically once every 16
seconds.  The functional form of the correction was provided by  the detector
development group at the University of California,  Berkeley, and its numerical
constants were determined from in-flight  calibration data.  It is applied by
the subroutine {\tt cf\_electronics\_dead\_time}.

A second contribution to the dead time comes from the way that the Instrument
Data System (IDS) processes counts coming from the detector. The IDS can 
accept at most 8000 counts per  second in time-tag mode and 32000
counts per second in histogram mode from the four detector segments (combined).
At higher count rates,  photon events are lost.  To correct for such losses, the
subroutine  {\tt cf\_ids\_dead\_time} compares the Active Image Counter (AIC)
count rate, measured at the back end of the detector electronics, with the
maximum allowed rate.  The IDS dead-time correction is the  ratio of these two
numbers (or unity, whichever is greater).

A third contribution occurs when time-tag data are
bundled into 64 kB data blocks in the IDS bulk memory.  This memory is
organized as a software FIFO (first-in, first-out) memory buffer, and the
maximum data transfer rate from it to the spacecraft recorder (the FIFO drain
rate) is approximately 3500 events per second.  At higher count rates, the
FIFO will eventually fill, resulting in the loss of one or  more data blocks.
The effect appears as a series of data drop-outs,  each a few seconds in length,
in the raw data files.  The correction,  computed by the subroutine
{\tt cf\_fifo\_dead\_time}, is simply the  ratio of the AIC count rate to the
FIFO drain rate.   When triggered, this correction incorporates (and replaces) 
the IDS correction discussed above.

The total dead-time correction (always $\geq 1.0$) is simply the product of the detector electronics and IDS corrections.  It is computed (by the subroutine {\tt cf\_apply\_dead\_time}) once each second and applied to the data by scaling the WEIGHT associated with each photon event.  The mean value of the detector electronics, IDS, and total dead-time corrections are stored in the DET\_DEAD, IDS\_DEAD, and TOT\_DEAD header keywords, respectively.  Other possible sources of dead time, such as losses due to the finite response time of the MCPs, have a much smaller effect and are ignored.

{\it Caveats:}  Our dead-time correction algorithms are inappropriate for very bright targets.  If the header keyword TOT\_DEAD $> 1.5$, then the exposure should not be considered photometric.  If the housekeeping file for a particular exposure is missing, the file header keywords from which the count rates are calculated appear to be corrupted, and either DET\_DEAD or IDS\_DEAD is $> 1.5$, then the dead-time correction is assumed to be meaningless and is set to unity.  In both of these cases, warning messages are written to the file header and the trailer file.

\subsubsection{Temperature-Dependent Changes in Detector Coordinates}\label{sec_stim}

The X and Y coordinates of a photon event do not correspond to a physical pixel on the detector, but are calculated from timing and voltage measurements of the incoming charge cloud \citep{Siegmund:97, Sahnow:SPIE:00}.  As a result, the detector coordinate system is subject to drifts in the detector electronics caused by temperature changes and other effects.  To track these drifts, two signals are periodically injected into the detector electronics. These ``stim pulses'' appear near the upper left and upper right corners of each detector, outside of the active spectral region. The stim pulses are well placed for tracking changes in the scale and offset of the X coordinate, but they are not well enough separated in Y to track scale changes along that axis.  The subroutine {\tt cf\_thermal\_distort} determines the X and Y centroids of the stim pulses, computes the linear transformation necessary to move them to their reference positions, and applies that transformation to the X and Y coordinates of each photon event in the regions of the stim pulses and in the active region of the detector.  Events falling within 64 pixels (in X and Y) of the expected stim-pulse positions are flagged by setting the stim-pulse bit in the LOC\_FLGS array (\S\ \ref{idf_format}).  In raw histogram files, the stim pulses are stored in a pair of image extensions.  If either of these extensions is missing, the pipeline reads the expected positions of the stim pulses from the calibration file STIM\_CAL and applies the corresponding correction.  This works (to first order) because the stim pulses drift slowly with time, though short-timescale variations cannot be corrected if the stim pulses are absent.

\subsubsection{Count-Rate Dependent Changes in Detector Y Scale}

For reasons not yet understood, the detector Y scale varies with the count rate, in the sense that the detector image for a high count-rate exposure is expanded in Y.  To measure this effect, we tabulated the positions of individual detector features (particularly bad-pixel regions) as a function of the FEC count rate (\S\ \ref{deadtime}) and determined the Y corrections necessary to shift each detector feature to its observed position in a low count-rate exposure.  From this information, we derived the calibration file RATE\_CAL for each detector segment.  The correction is stored as a two-dimensional image: the first dimension represents the count rate and the second is the observed Y pixel value.  The value of each image pixel is the Y shift (in pixels) necessary to move a photon to its corrected position.  The subroutine {\tt cf\_count\_rate\_y\_distort} applies this correction to each photon event in the active region of the detector.  For time-tag data, the FEC count rate is used to compute a time- and Y-dependent correction; for histogram data, the weighted mean of the FEC count rate is used to derive a set of shifts that depends only on Y.  

\subsubsection{Time-Dependent Changes in Detector Coordinates}\label{sec_tmxy}

As the detector and its electronics age, their properties change, resulting in small drifts in the computed coordinates of photon events.  These changes are most apparent in the Lyman $\beta$ airglow features observed in each of the three apertures of the LiF and SiC channels (\fig{detector1a}), which drift slowly apart in Y as the mission progresses, indicating a time-dependent stretch in the  detector Y scale.  To correct for this stretch, the subroutine {\tt cf\_time\_xy\_distort} applies a time-dependent correction (stored in the calibration file TMXY\_CAL) to the Y coordinate of each photon event in the active region of the detector.

{\it Caveats:}  Although there is likely to be a similar change to the X coordinate, no measurement of time-dependent drifts in that dimension is available, so no correction is applied.

\subsubsection{Geometric Distortion}\label{sec_geometric}

\begin{figure}
%\epsscale{.80}
\plotone{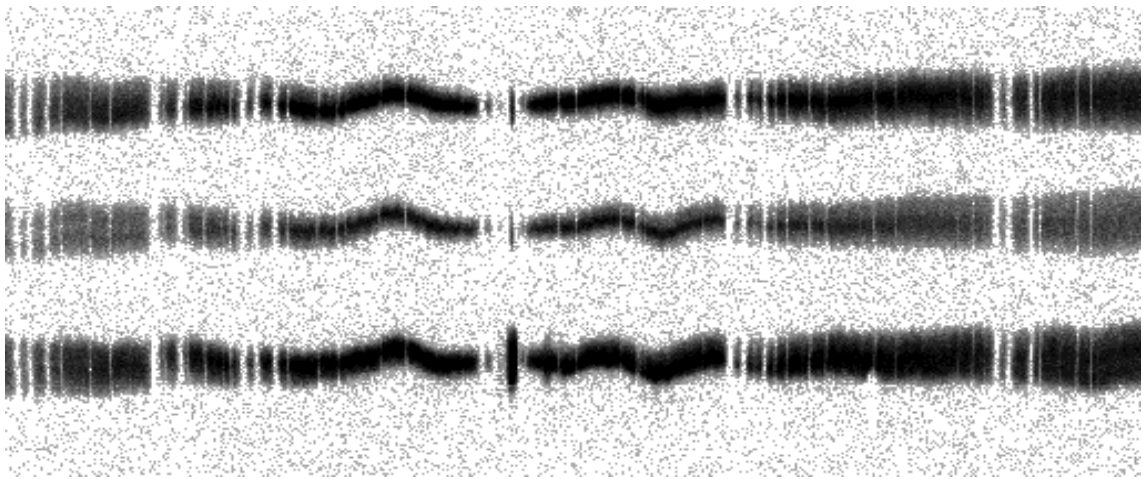}
\caption{Image of detector 1A in raw X and Y coordinates showing geometric distortion.  The image shows only a portion of the detector.  It was constructed from 3 separate exposures with stars in the HIRS, MDRS, and LWRS apertures of the LiF1A detector.}
\label{fig_geometric}
\end{figure}

In an image of the detector generated from raw X and Y coordinates, the spectrum is not straight, but wiggles in the Y dimension (\fig{fig_geometric}).  To map these geometric distortions, we made use of two wire grids (the so-called ``quantum efficiency'' and ``plasma'' grids) that lie in front of each detector segment.  Both grids are regularly spaced and cover the entire active area of the detectors.  Although designed to be invisible in the spectra, they cast sharp shadows on the detector when illuminated directly by on-board stimulation (or ``stim'') lamps.  We determined the shifts necessary to straighten these shadows.  Their spacing is approximately 1 mm, too great to measure fine-scale structure in the X dimension, but sufficient for the Y distortion.  Geometric distortions in the X dimension have the effect of compressing and expanding the spectrum in the dispersion direction, so the X distortion correction is derived in parallel with the wavelength calibration as described in \S\ \ref{sec_wavecal}.  The geometric distortion corrections are stored in a set of calibration files (GEOM\_CAL) as pairs of $16384 \times 1024$ images, one each for the X and Y corrections.  The value of each image pixel is the shift necessary to move a photon to its corrected position.  This shift is applied by the subroutine {\tt cf\_geometric\_distort}.

{\it Caveats:}  Though designed to be invisible, the wire grids can cast shadows that are visible in the spectra of astrophysical targets.  These shadows are the ``worms'' discussed in \S\ \ref{sec_worm}.

\subsubsection{Pulse-Height Variations in Detector X Scale}\label{sec_walk}

The \fuse\/ detectors convert each ultraviolet photon into a shower of
electrons, for which the detector electronics calculate the X and Y
coordinates and the total charge, or pulse height. Prolonged exposure to
photons causes the detectors to become less efficient at this  photon-to-electron
conversion (a phenomenon called ``gain sag''), and the mean pulse
height slowly decreases.  Unfortunately, the X coordinate of
low-pulse-height photon events is systematically miscalculated by the
detector electronics. As the pulse height decreases with time, 
spectral features appear to ``walk'' across the detector.
The strength of the effect depends on the cumulative photon exposure experienced
by each pixel and therefore varies with location on the detector.

To measure the error in X as a function of pulse height, we used data from long stim lamp exposures to construct a series of 32 detector images, each containing events with a single pulse height (allowed values range from 0 to 31).  We stepped through each image in X, computing the shift ($\Delta X$) necessary to align the shadow of each grid wire with the corresponding shadow in a standard image constructed from photon events with pulse heights between 16 and 20.  The shifts were smoothed to eliminate discontinuities and stored in calibration files (PHAX\_CAL) as a two-dimensional image: the first dimension represents the observed X coordinate, and the second is the pulse height.  The value of each image pixel is the walk correction ($\Delta X$) to be added to the observed value of X.  This correction, assumed to be independent of detector Y position, is applied by the subroutine {\tt cf\_pha\_x\_distort}.

{\it Caveats:} For time-tag data, the walk correction is straightforward and reasonably accurate.  For histogram data, pulse-height information is unavailable, so the subroutine {\tt cf\_modify\_hist\_pha} assigns to each photon event the mean pulse height for that aperture, derived from contemporaneous time-tag observations and stored in the calibration file PHAH\_CAL. While this trick places time-tag and histogram data on the same overall wavelength scale, small-scale coordinate errors due to localized regions of gain sag (\eg, around bright airglow lines, particularly Lyman $\beta$) remain uncorrected in histogram spectra. 

\subsubsection{Detector Active Region}\label{sec_active_region}

When the IDF is first created, photon events with coordinates outside the active region of the detector are flagged as bad (\S\ \ref{init}).  Once their coordinates are converted to the FARF, the subroutine {\tt cf\_active\_region} flags as bad any photons that have been repositioned  beyond the active region of the detector.  These limits are read from the electronics calibration file (stored under the header keyword ELEC\_CAL).   Allowed values are $800 \leq {\rm XFARF} \leq 15583$, $0 \leq {\rm YFARF} \leq 1023$.  The active-area bit is written to the LOC\_FLGS array.

\subsubsection{Uncorrected Detector Effects}\label{sec_uncorrected}

CalFUSE does not perform any sort of flat-field correction.  Pre-launch flat-field observations differ sufficiently from in-flight data to make them unsuitable for this purpose, and in-flight flat-field data are unavailable.  (Even if such data were available, any flat-field correction would be only approximate, because MCPs do not have physical pixels for which pixel-to-pixel variations can be clearly delineated; \S\ \ref{sec_stim}).  As a result, detector fixed-pattern noise limits the signal-to-noise ratio achievable in observations of bright targets.  To the extent that grating, mirror, and spacecraft motions shift the spectrum on the detector during an exposure, these fixed-pattern features may be averaged out.  For some targets, we deliberately move the FPAs between exposures to place the spectrum on different regions of the detector.  Combining the extracted spectra from these exposures can significantly improve the resulting signal-to-noise ratio (\S\ \ref{sec_fpa_motion}).  Other detector effects (including the moir\'{e} pattern discussed in \S\ \ref{sec_moire}) are described in the \anchor{http://archive.stsci.edu/fuse/dhbook.html}{{\em FUSE Instrument and Data Handbook.}}

\subsection{Screen Photons}\label{sec_screen}

The module {\tt cf\_screen\_photons} calls subroutines designed to identify periods of potentially bad data, such as Earth limb-angle violations, SAA passages, and detector bursts.  A distinct advantage of CalFUSE v3 over earlier versions of the pipeline is that bad data are not discarded, but merely flagged, allowing users to modify their selection criteria without having to re-process the data.  To speed up processing, the pipeline calculates the various screening parameters once per second throughout the exposure, sets the corresponding flags in the STATUS\_FLAGS array of the timeline table, then copies the flags from the appropriate entry of the timeline table into the TIMEFLGS array for each photon event (\S\ \ref{idf_format}).  Many of the screening parameters applied by the pipeline are set in the screening parameter file (SCRN\_CAL).  Other parameters are stored in various calibration files as described below.

\subsubsection{Airglow Events}

Numerous geocoronal emission features lie within the \fuse\/ waveband \citep{Feldman:01}.
While the pipeline processes airglow photons in the same manner as all other photon events in the target aperture, it is occasionally helpful to exclude from consideration regions of the detector likely to be contaminated by geocoronal or scattered solar emission.  These regions are listed in the calibration file AIRG\_CAL; the subroutine {\tt cf\_screen\_airglow} flags as airglow (by setting the airglow bit of the LOC\_FLGS array in the photon-event list) all photon events falling within the tabulated regions -- even for data obtained during orbital night, when many airglow features are absent.

\subsubsection{Limb Angle}

Spectra obtained when a target lies near the earth's limb are contaminated by scattered light from strong geocoronal Lyman $\alpha$ and \oone\ emission.  To minimize this effect, the subroutine {\tt cf\_screen\_limb\_angle} reads the LIMB\_ANGLE array of the timeline table, identifies periods when the target violates the limb-angle constraint, and sets the corresponding flag in the STATUS\_FLAGS array of the timeline table.  Minimum limb angles for day and night observations are read from the BRITLIMB and DARKLIMB keywords of the screening parameter file and copied to the IDF file header.  The default limits are 15\degr\ during orbital day and 10\degr\ during orbital night.

\subsubsection{SAA Passages}

The South Atlantic Anomaly (SAA) marks a depression in the earth's magnetic field that allows particles trapped in the Van Allen belts to reach low altitudes.  The high particle flux in this region raises the background count rate of the \fuse\/ detectors to unacceptable levels.  The subroutine {\tt cf\_screen\_saa} compares the spacecraft's ground track, recorded in the LONGITUDE and LATITUDE arrays of the timeline table, with the limits of the SAA (stored in the calibration file SAAC\_CAL as a binary table of latitude-longitude pairs) and flags as bad periods when data were taken within the SAA.  Our SAA model was derived from orbital information and on-board counter data from the first three years of the \fuse\/ mission.

{\it Caveats:}  Because the SAA particle flux is great enough to damage the \fuse\/ detectors, we end most exposures before entering the SAA and lower the detector high voltage during each SAA pass.  As a result, very little data is actually flagged by this screening step.

\subsubsection{Detector High Voltage}\label{sec_hv}

The detector high voltage is set independently for each detector  
segment (1A, 1B, 2A, 2B).  During normal operations, the voltage
on each segment alternates between its nominal full-voltage and a reduced SAA
level. The SAA level is low enough that the detectors are
not damaged by the high count rates that result from SAA passes, 
and it is often used between science exposures
to minimize detector exposure to bright airglow emission. The
full-voltage level is the normal operating voltage used during science exposures. 
It is raised regularly to compensate for the effects of detector gain sag.   
Without this compensation, the mean pulse height of real photon  
events would gradually fall below our detection threshold.   
Unfortunately, there is a limit above which the full-voltage level  
cannot be raised.  Detector segment 2A reached this limit in 2003 and has
not been raised since; it will gradually become less sensitive as the fraction of low-pulse-height events increases.  The subroutine {\tt cf\_screen\_high\_voltage} reads the instantaneous value of the detector high voltage from the HIGH\_VOLTAGE array of the timeline table, compares it with the nominal full-voltage level (stored as a function of time in the calibration file VOLT\_CAL), and flags periods of low voltage as bad.

For any number of reasons, an exposure may be obtained with the detector high voltage at less than the full-voltage level.  To preserve as much of this data as possible, we examined all of the low-voltage exposures taken during the first four years of the mission and found that, for detector segments 1A, 1B, and 2B, the data quality is good whenever the detector high voltage is greater than 85\% of the nominal (time-dependent) full-voltage level.  For segment 2A, data obtained with the high voltage greater than 90\% of full are good, lower than 80\% are bad, and between 80 and 90\% are of variable quality.  In this regime, the pipeline flags the affected data as good, but writes warning messages to both the IDF header and the trailer file.  When this warning is present in time-tag data, the user should examine the distribution of pulse heights in the target aperture to ensure that the photon events are well separated from the background (\S\ \ref{sec_pha}).  For histogram data, the spectral photometry and wavelength scale are most likely to be affected.

{\it Caveats:}  If the header keywords indicate that the detector voltage was high, low, or changed during an exposure, the IDF initialization routines (\S\ \ref{init}) write a warning message to the trailer file.  If a valid housekeeping file is available for the exposure, this warning may be safely ignored, because the pipeline uses housekeeping information to populate the HIGH\_VOLTAGE array in the timeline table and properly excludes time intervals when the voltage was low.  If the housekeeping file is not present, each entry of the HIGH\_VOLTAGE array is set to the ``HV bias maximum setting'' reported in the IDF header.  In this case, the pipeline has no information about time-dependent changes in the detector high voltage, and warnings about voltage-level changes should be investigated by the user.

\subsubsection{Event Bursts}\label{screen_burst}

\begin{figure}
\epsscale{.80}
\plotone{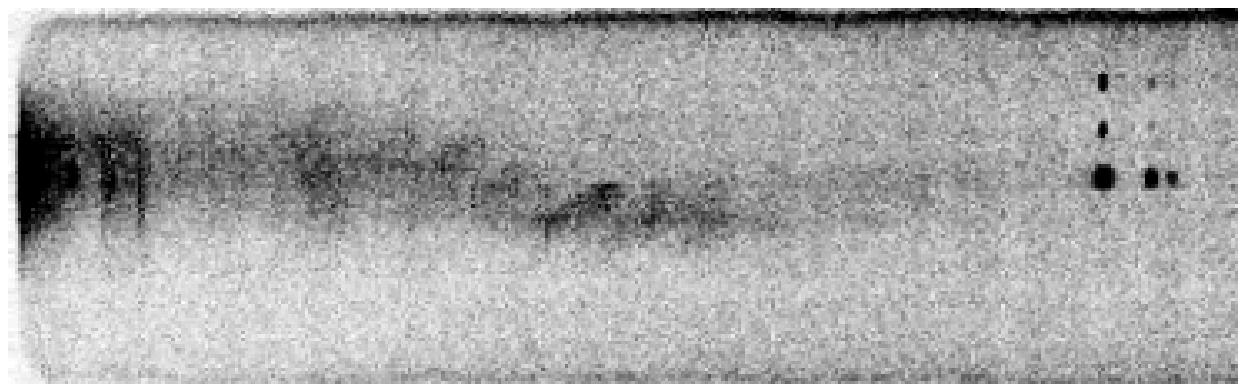}
\plotone{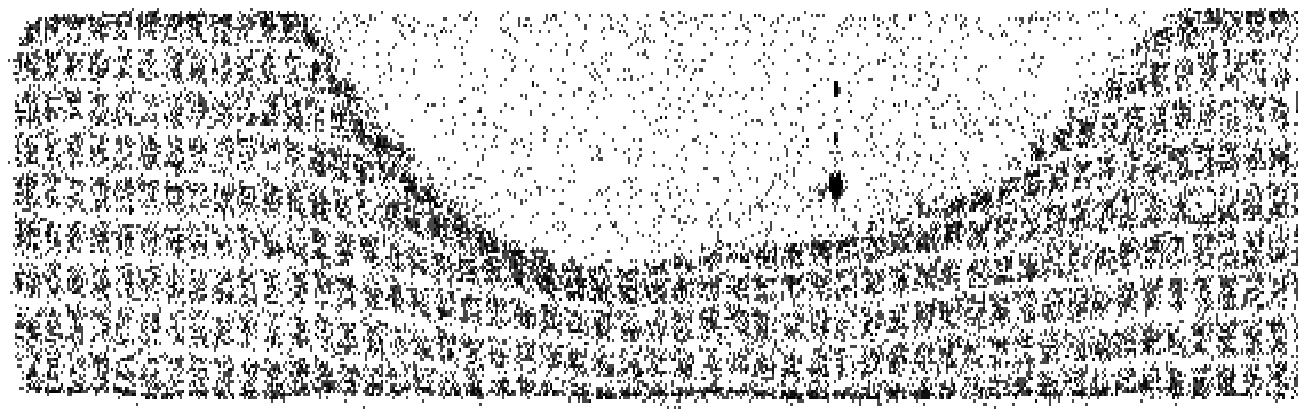}
\caption{Segments of detector 1A images showing filamentary (top) and checkerboard (bottom) bursts.  Checkerboard bursts typically fill the detector, save for the region around the LiF Lyman $\beta$ lines on detector 1A.}
\label{burst}
\end{figure}

Occasionally, the \fuse\/ detectors register large count rates for
short periods of time. These event bursts can occur on one or more
detectors and often have a complex distribution across the detector, including scalloping and
sharp edges (\fig{burst}). CalFUSE includes a module that screens the data to identify and exclude bursts.  The subroutine {\tt cf\_screen\_burst} computes the time-dependent count rate using data from background regions of the detector (excluding airglow features) and applies a median filter to reject time intervals whose count rates differ by more than 5 standard deviations (the value may be set by the user) from the mean. The algorithm rejects any time interval in which the background rate rises rapidly, as when an exposure extends into an SAA or the target nears the earth limb.  The background rate computed by the burst-rejection algorithm is stored in the BKGD\_CNT\_RATE array of the timeline table and included on the count-rate plots generated for each exposure (\S\ \ref{sec_trailer}).  Burst rejection is possible only for data obtained in time-tag mode.

{\it Caveats:}  Careful examination of long background observations reveals that many are contaminated by emission from bursts too faint to trigger the burst-detection algorithm.  Observers studying, for example, diffuse emission from the interstellar medium should be alert to the possibility of such contamination. 

\subsubsection{Spacecraft Drift}\label{sec_flag_jitter}

Pointing of the \fuse\/ spacecraft was originally controlled with four reaction wheels, which typically maintained a pointing accuracy of 0.2--0.3 arc seconds. In late 2001, two of the reaction wheels failed, and it became necessary to control the spacecraft orientation along one axis with magnetic torquer bars. The torquer bars can exert only about 10\% of the force produced by the reaction wheels, and the available force depends on the strength and relative orientation of the earth's magnetic field. Thus, spacecraft drift increased dramatically along this axis, termed the antisymmetric or A axis. 
Drifts about the A axis shift the spectra of point-source targets in a direction 45\degr\ from the dispersion direction (\ie, $\Delta X = \Delta Y$).  
These motions can substantially degrade the resolution of the spectra, so procedures have been implemented to correct the data for spacecraft motion during an exposure.  For time-tag observations of point sources, we reposition individual photon events.  For histogram observations, we correct only for the exposure time lost to large excursions of the spacecraft.  The ability to correct for spacecraft drift became even more important when a third reaction wheel failed in 2004 December.

The correction of photon-event coordinates for spacecraft motion is discussed in \S\ \ref{sec_correct_jitter}.  During screening, the subroutine {\tt cf\_screen\_jitter} merely flags times when the target is out of the aperture, defined as those for which either $\Delta X$ or $\Delta Y$, the pointing error in the dispersion or cross-dispersion direction, respectively, is greater than 30\arcsec, the width of the LWRS aperture.  These limits, set by the keywords DX\_MAX and DY\_MAX in the CalFUSE parameter file (PARM\_CAL), underestimate the time lost to pointing excursions, but smaller limits can lead to the rejection of good data for some channels.  Also flagged as bad are times when the jitter tracking flag TRKFLG $= -1$, indicating that the spacecraft is not tracking properly.  If TRKFLG $ = 0$, no tracking information is available and no times are flagged as bad.  Pointing information is read from the jitter file (JITR\_CAL; \S\ \ref{sec_jitter}).  If the jitter file is not present or the header keyword JIT\_STAT $= 1$ (indicating that the jitter file is corrupted), {\tt cf\_screen\_jitter} issues a warning and exits; again, no times are flagged as bad.

\subsubsection{User-Defined Good-Time Intervals}\label{sec_user_defined}

One bit of the status array is reserved for user-defined GTIs.  For example, to extract data corresponding to a particular phase of a binary star orbit, one would flag data from all other phases as bad.  A number of tools exist to set this flag, including {\tt cf\_edit} (available from MAST).  CalFUSE users may specify good-time intervals by setting the appropriate keywords (NUSERGTI, GTIBEG01, GTIEND01, \etc) in the screening parameter file.  (Times are in seconds from the exposure start time.)  If these keywords are set, the subroutine {\tt cf\_set\_user\_gtis} flags times outside of these good-time intervals as bad. 

\subsubsection{Time-Dependent Status Flags}\label{sec_photon_flags}

Once the status flags in the timeline table are populated, the subroutine {\tt cf\_set\_photon\_flags} copies them to the corresponding entries in the photon event list.  For time-tag data, this process is straightforward: match the times and copy the flags.  Header keywords in the IDF record the number of photon events falling in bad time intervals or outside of the detector active area; the number of seconds lost to bursts, SAAs, \etc; and the remaining night exposure time.  If more than 90\% of the exposure is lost to a single cause, an explanatory note is written to the trailer file.

The task is more difficult for histogram data, for which photon-arrival information is unavailable.  We distinguish between time flags that represent periods of lost exposure time (low detector voltage or target out of aperture) and those that represent periods of data contamination (limb angle violations or SAAs).  For the former, we need only modify the exposure time; for the latter, we must flag the exposure as being contaminated.  Our goal is to set the individual photon flags and header keywords so that the pipeline behaves in the following way:  When processing a single exposure, it treats all photon events as good.  When combining data from multiple exposures, it excludes contaminated exposures (defined below).  To this end, we generate an 8-bit status word containing only day/night information: if the exposure is more than 10\% day, the day bit is set.  This status word is copied onto the time-dependent status flag of each photon event.  We generate a second 8-bit status word containing information about limb-angle violations and SAAs: if a single second is lost to one of these events, the corresponding flag is set and a message is written to the trailer file.  (To avoid rejecting an exposure that, for example, abuts an SAA, we ignore its initial and final 20 seconds in this analysis.)  The status word is stored in the file header keyword EXP\_STAT (unless that keyword has already been set; see \S\ \ref{init} and \S\ \ref{sec_digitizer}).  The routines used by the pipeline to combine data from multiple exposures into a single spectrum (\S\ \ref{obs_files}) reject data files in which this keyword is non-zero.  The number of bad events, the exposure time lost to periods of low detector voltage or spacecraft jitter, and the exposure time during orbital night are written to the file header, just as for time-tag data.

Only in this subroutine is the DAYNIGHT keyword read from the screening parameter file and written to the IDF file header.  Allowed values are DAY, NIGHT, and BOTH.  The default is BOTH.  For most flags, if the bit is set to 1, the photon event is rejected.  The day/night flag is different: it is always 1 for day and 0 for night.  The pipeline must read and interpret the DAYNIGHT keyword before accepting or rejecting an event based on the value of its day/night flag.

\subsubsection{Good-Time Intervals}\label{sec_GTI}

Once the time-dependent screening is complete, the subroutine {\tt cf\_set\_good\_time\_intervals} calculates a new set of good-time intervals from information in the timeline table and writes them to the second extension of the IDF (\S\ \ref{init}).  For time-tag data, all of the TIMEFLGS bits are used and the DAYNIGHT filter is applied.  For histogram data, the bits corresponding to limb-angle violations and SAAs are ignored, since data arriving during these events cannot be excluded.  The DAYNIGHT filter is applied (assuming that all are day photons if the exposure is more than 10\% day).  The exposure time, EXPTIME = $\Sigma$ (STOP$ - $START), summed over all entries in the GTI table, is then written to the IDF file header.

\subsubsection{Histogram Arrival Times}\label{sec_hist_times}

For histogram data, all of the photon events in an IDF are initially assigned an arrival time equal to the midpoint of the exposure.  Should this instant fall in a bad-time interval, the data may be rejected by a subsequent step of the pipeline or one of our post-processing tools.  To avoid this possibility, the subroutine {\tt cf\_modify\_hist\_times} resets all photon-arrival times to the midpoint of the exposure's longest good-time interval.  This subroutine is not called for time-tag data.  

\subsubsection{Bad-Pixel Regions}\label{sec_flag_pot_holes}

Images of the \fuse\/ detectors reveal a number of dead spots that may be surrounded by a bright ring (see the \anchor{http://archive.stsci.edu/fuse/dhbook.html#DEAD}{{\em FUSE Instrument and Data Handbook}}\/ for examples).  The subroutine {\tt cf\_screen\_bad\_pixels} reads a list of bad-pixel regions from a calibration file (QUAL\_CAL) and flags as bad all photon events whose XFARF and YFARF coordinates fall within the tabulated limits.  A bad-pixel map, constructed later in the pipeline (\S\ \ref{sec_bpm}), is used by the optimal-extraction algorithm to correct for flux lost to dead spots.

\subsubsection{Pulse Height Limits}\label{sec_pha}

For time-tag data, the pulse height of each photon event is recorded in the IDF.  Values range from 0 to 31 in arbitrary units.  
A typical pulse-height distribution has a peak at low values due to the intrinsic detector background, a Gaussian-like peak near the middle of the range due to ``real'' photons, and a tail of high pulse-height events, which likely represent the superposition of two photons and therefore are not reliable.  
In addition, the detector electronics selectively discard high pulse-height events near the top and bottom of the detectors (\ie, with large or small values of Y).
We can thus improve the signal-to-noise ratio of faint targets by rejecting photon events with extreme pulse-height values.  
Pulse-height limits (roughly 2--24) are defined for each detector segment in the screening parameter file (SCRN\_CAL).  
The subroutine {\tt cf\_screen\_pulse\_height} flags photon events with pulse heights outside of this range (by setting the appropriate bit in the LOC\_FLGS array; \S\ \ref{idf_format}) and writes the pulse-height limits used and the number of photon events rejected to the IDF file header.  This procedure is not performed on histogram data.

{\it Caveats:} We do not recommend the use of narrow pulse-height ranges to reduce the detector background in \fuse\/ data. Careful analysis has shown that limits more stringent than the default values can result in significant flux losses across small regions of the detector, particularly in the LiF1B channel, resulting in apparent absorption features that are not real.

\subsection{Remove Motions}\label{remove_motions}

Having corrected the data for various detector effects and identified periods of bad data, we continue to work backwards through the instrument, correcting for spectral motions on the detector due to the movements of various optical components -- and even of the spacecraft itself.  This task is performed by the module {\tt cf\_remove\_motions}.  It begins by reading the XFARF and YFARF coordinates of each photon event from the IDF.  It concludes by writing the motion-corrected coordinates to the X and Y arrays of the same file.

\subsubsection{Locate Spectra on the Detector}\label{sec_find_spectra}

\begin{figure}
%\epsscale{.80}
\plotone{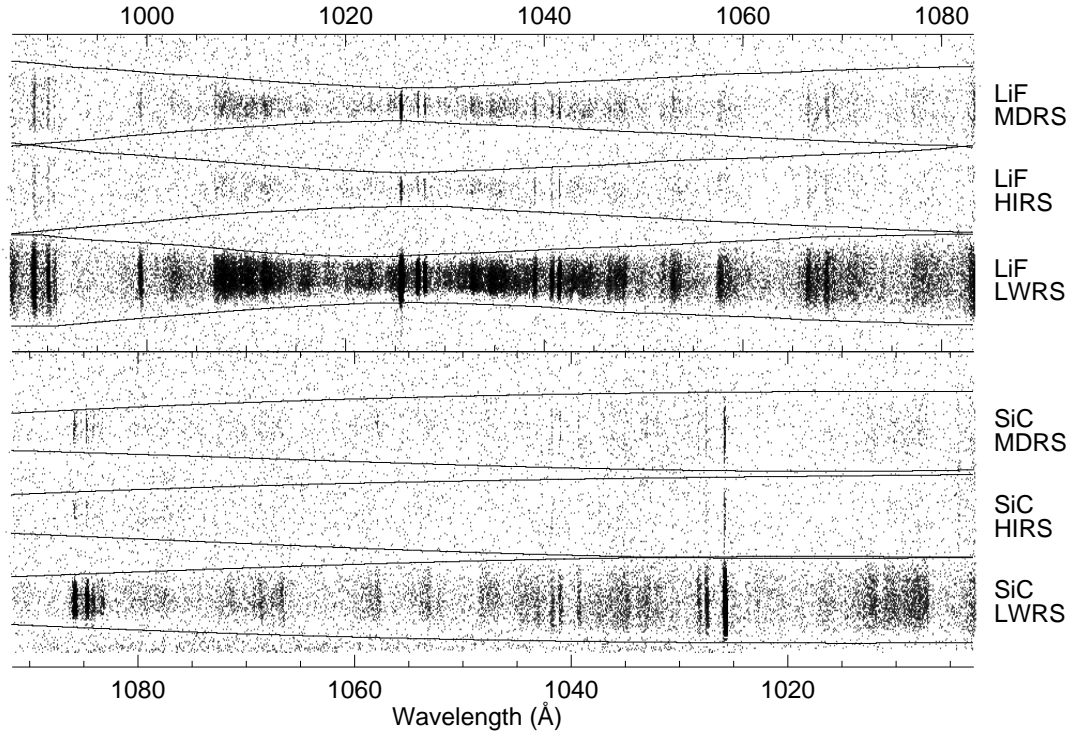}
\caption{Image of detector segment 1A during a bright-earth observation.  All lines are geocoronal.  Note the strong Lyman $\beta$ (1026 \AA) feature in each spectrum.  The data have been fully corrected for detector and other distortions.  Extended-source extraction windows for all three apertures in both the LiF and SiC channels are marked; point-source extraction windows are somewhat narrower in Y.  Instrumental astigmatism is responsible for the bow-tie shape of each spectrum.  The region shown corresponds to detector pixels 900 to 15,300 in X and 0 to 915 in Y.}
\label{detector1a}
\end{figure}

The LiF and SiC channels each produce three spectra, one from each aperture, for a total of six spectra per detector segment (\fig{detector1a}).  Because motions of the optical components can shift these spectra on the detector, the first step is to determine the Y centroid of each. To do this, we use the following algorithm: First, we project the airglow photons onto the Y axis (summing all values of X for each value of Y) and search the resulting histogram for peaks within 70 pixels of the expected Y position of the LWRS spectrum.  If the airglow feature is sufficiently bright (33 counts in 141 Y pixels), we adopt its centroid as the airglow centroid for the LWRS aperture and compute its offset from the expected value stored in the CHID\_CAL calibration file.  If the airglow feature is too faint, we adopt the expected centroid and assume an offset of zero.  We apply the offset to the expected centroids of the MDRS and HIRS apertures to obtain their airglow centroids.  Second, we project the non-airglow photons onto the Y axis and subtract a uniform background.  Airglow features fill the aperture, but point-source spectra are considerably narrower in Y and need not be centered in the aperture.  For each aperture, we search for a 5$\sigma$ peak within 40 pixels of the airglow centroid.  If we find it, we use its centroid; otherwise, we use the airglow centroid.  This scheme, implemented in the subroutine {\tt cf\_find\_spectra}, allows for the possibility that an astrophysical spectrum may appear in a non-target aperture.

For each of the six spectra, two keywords are written to the IDF file header: YCENT contains the computed Y centroid, and YQUAL contains a quality flag.  The flag is HIGH if the centroid was computed from a point-source spectrum, MEDIUM if computed from an airglow spectrum, and LOW if the tabulated centroid was used.  It is possible for the user to specify the target centroid by setting the SPEX\_SIC and SPEC\_LIF keywords in the CalFUSE parameter file (PARM\_CAL).  Two other keywords, EMAX\_SIC and EMAX\_LIF, limit the offset between the expected and calculated centroids: if the calculated centroid differs from the predicted value by more than this limit, the pipeline uses the default centroid. 

{\it Caveats:} On detector 1, the SiC LWRS spectrum falls near the bottom edge of the detector (\fig{detector1a}).  Because the background level rises steeply near this edge, the calculated centroid can be pulled (incorrectly) to lower values of Y, especially for faint targets.

\subsubsection{Assign Photons to Channels}\label{sec_assign_photon}

The subroutine {\tt cf\_identify\_channel} assigns each photon to a channel, where ``channel'' now refers to one of the six spectra on each detector (\fig{detector1a}).  For each channel, extraction windows for both point-source and extended targets are tabulated in the calibration file CHID\_CAL along with their corresponding spectral Y centroids.  These extraction limits encompass at least 99.5\% of the target flux.  For the target channels, identified in the APERTURE header keyword, we use either the point-source or extended extraction windows, as indicated by the SRC\_TYPE keyword; for the other (presumably airglow) channels, we use the extended extraction windows.  The offset between the calculated and tabulated spectral Y centroids (\S\ \ref{sec_find_spectra}) is used to shift each extraction window to match the data.

To insure that, should two extraction windows overlap, photon events falling in the overlap region are assigned to the more likely channel, photon coordinates (XFARF and YFARF) are compared with the extraction limits of the six spectral channels in the following order:  first the target channels (LiF and SiC); then the airglow channels (LiF and SiC) corresponding to the larger non-target aperture; and finally the airglow channels (LiF and SiC) corresponding to the smaller non-target aperture.  For example, if the target were in the MDRS aperture, the search order would be MDRS LiF, MDRS SiC, LWRS LiF, LWRS SiC, HIRS LiF, and HIRS SiC.  The process stops when a match is made.  The channel assignment of each photon event is stored in the CHANNEL array (\S\ \ref{idf_format}); photon events that do not fall in an extraction window are assigned a CHANNEL value of 0.

Channel assignment is performed twice, once before the motion corrections and once after.  The first time, all extraction windows are padded by $\pm 10$ Y pixels to accommodate errors in the channel centroids.  The second time, no padding is applied to time-tag data.  Histogram data, which are generally binned by 8 pixels in Y, present a special challenge: The geometric correction (\S\ \ref{sec_geometric}) can move a row of data out of the extraction window, producing a significant loss of flux.  To prevent this, histogram extraction windows are padded by an additional $\pm 8$ Y pixels (or an amount equal to the binning factor in Y, if other than 8).

\subsubsection{Track Y Centroids with Time}

For the LiF and SiC target apertures, the subroutine {\tt cf\_calculate\_ycent\_motion} computes the spectral Y centroid as a function of time throughout the exposure.  The algorithm requires 500 photon events to compute an average, so the centroid is updated more often for bright targets than for faint ones.  Photon events flagged as airglow are ignored.  The results are stored in the YCENT\_LIF and YCENT\_SIC arrays of the timeline table, but are not currently used by the pipeline.  This calculation is not performed for data obtained in histogram mode.

\subsubsection{Correct for Grating Motion}\label{sec_grating_motion}

The \fuse\/ spectrograph gratings are subject to small, thermally-induced motions on orbital, diurnal, and precessional (60-day) timescales; an additional long-term, non-periodic drift is also apparent.  These motions can shift the target and airglow spectra by as much as 15 pixels (peak to peak) in both the X and Y dimensions.  Measurements of the Lyman $\beta$ airglow line in thousands of exposures obtained throughout the mission reveal that the gratings' orbital motion depends on three parameters: beta angle (the angle between the target and the anti-sun vector), pole angle (the angle between the target and the orbit pole), and spacecraft roll angle (east of north, stored in the file-header keyword APER\_PA). The subroutine {\tt cf\_grating\_motion} compares the beta, pole, and roll angles of the spacecraft with a grid of values in the calibration file GRAT\_CAL, reads the appropriate correction, and computes the X and Y photon shifts.  The grating-motion correction is applied to all photon events with CHANNEL $> 0$; photon events not assigned to a channel are not moved.

{\it Caveats:}  Some combinations of beta and pole angle are too poorly sampled for us to derive a grating-motion correction; for these regions, no correction is applied.  At present, only corrections for the orbital and long-term grating motions are available.  Because all photon events in histogram data are assigned the same arrival time (the midpoint of the longest good-time interval), they receive the same grating-motion correction.

\subsubsection{Correct for FPA Motion}\label{sec_fpa_motion}

The four focal-plane assemblies (shown in \fig{toaster}) can be moved independently in either the X or Z direction.  FPA motions in the X direction are used to correct for mirror misalignments and to perform FP splits (described below).  
FPA motions in the Z direction are used to place the apertures in the focal plane of the spectrograph.  (Strictly speaking, an FPA moves along the tangent to or the radius of the spectrograph Rowland circle, not the X and Z axes shown in \fig{toaster}.)  Both motions change the spectrograph entrance angle, shifting the target spectrum on the detector.
The \fuse\/ wavelength calibration is derived from a single stellar observation obtained at a particular FPA position.  The subroutine {\tt cf\_fpa\_position} computes the shift in pixels ($\Delta X$) necessary to move each spectrum from its observed X position on the detector to that of the wavelength-calibration target.
The X and Z positions of the LiF and SiC FPAs are stored in file header keywords, various spectrograph parameters are stored in a calibration file (SPEC\_CAL), and the wavelength calibration and the FPA position of the wavelength-calibration target are stored in the WAVE\_CAL file. 
Shifts are computed for both the LiF and SiC channels; the appropriate shift is applied to all photon events with CHANNEL $> 0$; photon events not assigned to a channel are ignored.

The \fuse\/ detectors suffer from fixed-pattern noise.  Astigmatism in the instrument spreads a typical resolution element over several hundred detector pixels (predominantly in the cross-dispersion dimension), mitigating this effect, but to achieve a signal-to-noise ratio greater than $\sim$ 30, one must remove the remaining fixed-pattern noise.  A useful technique is the focal-plane split.  FP splits are performed by obtaining a series of MDRS or HIRS exposures
at several FPA X positions.  Moving the FPA in the X dimension (and moving the satellite to center the target in the aperture) between exposures shifts both target and airglow spectra in the dispersion direction on the detector.  CalFUSE shifts each spectrum to the standard X position expected by our wavelength calibration routines.  If the signal-to-noise ratio in the spectra obtained at each FPA position is
high enough, it is possible for the user to disentangle the source spectrum from the detector fixed-pattern noise; however, simply combining extracted spectra obtained at different FPA positions will average out most of the small-scale detector features.

\subsubsection{Correct for Mirror Motion}\label{sec_mirror}

The spectrograph mirrors are subject to thermal motions that shift the target's image within the spectrograph aperture and thus its spectrum in both X and Y on the detector.  A source in either of the SiC channels may move as much as 6\arcsec\ in a period of 2 ks.  This motion has two effects on the data: first, flux will be lost if the source drifts (partially or completely) out of the aperture; second, spectral resolution will be degraded (for LWRS observations) as the spectrum shifts on the detector.  Diffuse sources, such as airglow emission, fill the aperture, so their spectra are unaffected by mirror motion.

When the LiF1 channel is used for guiding, motions of the LiF1 mirror are corrected by the spacecraft itself.  Only the LiF2 and SiC spectra must be corrected by CalFUSE.  In theory, the switch from LiF1 to LiF2 as the primary channel for guiding the spacecraft (\S\ \ref{sec_fuse}) should require another set of calibration files.  In practice, the LiF2 mirror motion in the dispersion direction tracks that of the LiF1 mirror.  The mirror-motion correction is stored as a function of time since orbital sunset (via the TIME\_SUNSET array in the timeline table) in the calibration file MIRR\_CAL.  The correction ($\Delta X$) is applied by the subroutine {\tt cf\_mirror\_motion} to all photon events within the target aperture; photon events in other apertures and those not assigned to a channel are ignored.  This correction is not applied to extended sources.  Because all photon events in histogram data are assigned the same arrival time (generally the midpoint of the exposure), they receive the same mirror-motion correction.

{\it Caveats:}   We correct only the relative mirror motion during an orbit, not the absolute mirror offset based on longer-term trends.  We do not correct for mirror motions in the Y dimension.  Finally, because the shifts for the SiC1 and SiC2 mirrors are similar, we adopt a single correction for both channels.  

\subsubsection{Correct for Spacecraft Motion}\label{sec_correct_jitter}

Spacecraft motions during an exposure shift the target spectrum on the detector and thus degrade spectral resolution.  The subroutine {\tt cf\_satellite\_jitter} uses pointing information stored in the jitter file (JITR\_CAL; \S\ \ref{sec_jitter}) to correct the observed coordinates of photon events for these motions.  Pointing errors in arc seconds are converted to X and Y pixel shifts and applied to all photon events within the target aperture; events in other apertures and those not assigned to a channel are ignored.  The correction is applied only if the jitter tracking flag TRKFLG $ > 0$, indicating that valid tracking information is available.  TRKFLG values rise from 1 to 5 as the reliability of the pointing information increases.  The minimum acceptable value of the TRKFLG may be adjusted by modifying the TRKFLG keyword in the CalFUSE parameter file (PARM\_CAL).

\subsubsection{Recompute Spectral Centroids}

Once all spectral motions are removed from the data, the subroutine {\tt cf\_calculate\_y\_centroid} recomputes the spectral Y centroids.  Separate source and airglow centroids are determined for each aperture in turn, from largest to smallest.  (The former is meaningless if the aperture does not contain a source.)  The offset between the measured airglow centroid in the LWRS aperture and the tabulated centroid (from the calibration file CHID\_CAL) is used to compute the airglow centroids for the MDRS and HIRS apertures; the computed MDRS and HIRS airglow centroids are ignored.  The YCENT value written to the IDF file header is determined by the quality flag previously set by {\tt cf\_find\_spectra} (\S\ \ref{sec_find_spectra}): if YQUAL = HIGH, the source centroid is used; if YQUAL = MEDIUM, the airglow centroid is used; and if YQUAL = LOW, the tabulated centroid is used.  The SPEX\_SIC, SPEC\_LIF, EMAX\_SIC, and EMAX\_LIF keywords in the CalFUSE parameter file (PARM\_CAL) have the effects discussed in \S\ \ref{sec_find_spectra}.

\subsubsection{Final Assignment of Photons to Channels}

The final assignment of each photon event to a channel is performed by {\tt cf\_identify\_channel}, just as in \S\ \ref{sec_assign_photon}, but with two modifications:  First, we consider only photon events with CHANNEL $>$ 0; unassigned events (which are not motion corrected) remain unassigned.  Second, we do not pad the extraction windows by an additional $\pm 10$ pixels in Y, though the extraction windows for histogram data are padded by $\pm 8$ Y pixels  (or an amount equal to the binning factor in Y, if other than 8), as before.

\subsubsection{Compute Count Rates}

For time-tag data, {\tt cf\_target\_count\_rate} computes the count rate in the target aperture for the LiF and SiC channels.  To account for dead-time effects, the contents of the WEIGHT array are used.  Events in airglow regions are excluded, but no other filters are applied to the data.  Results are written to the LIF\_CNT\_RATE and SIC\_CNT\_RATE arrays of the timeline table.  For histogram data, the initial values of these arrays, taken from the housekeeping file (\S\ \ref{idf_format}), are scaled by the value of the header keyword DET\_DEAD.

\subsection{Wavelength Calibration}\label{wavelength}

Once converted to the FARF and corrected for optical and spacecraft drifts, the data can be wavelength calibrated.  The module {\tt cf\_assign\_wavelength} performs three tasks: first, it corrects for astigmatism in the spectrograph optics; second, it applies a wavelength calibration to each photon event; and third, it shifts the wavelengths to a heliocentric reference frame. 
The derivation of the \fuse\/ wavelength scale is discussed in \S\ \ref{sec_wavecal}.

\subsubsection{Astigmatism Correction}

\begin{figure}
\epsscale{.80}
\plotone{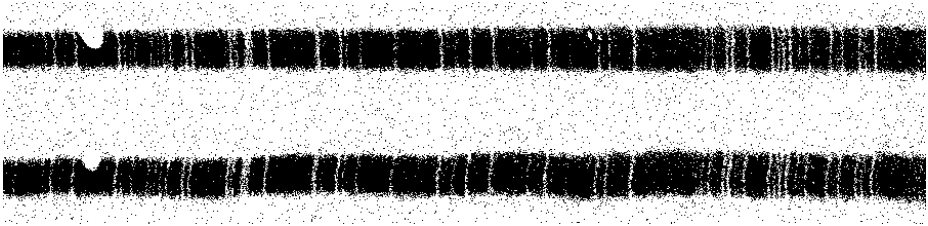}
\caption{Segment of LiF1A spectrum before (bottom) and after (top) astigmatism correction.  Note the reduction of curvature in the absorption features.  A detector dead spot is present on the left side of the figure.}
\label{astig}
\end{figure}

The astigmatic height of \fuse\/ spectra perpendicular to the dispersion axis is significant and varies as a function of wavelength (\fig{detector1a}).  Moreover, spectral features show considerable curvature, especially near the ends of the detectors where the astigmatism is greatest. The subroutine {\tt cf\_astigmatism} shifts each photon event in X to correct for the spectral-line curvature introduced by the \fuse\/ optics, providing a noticeable improvement in spectral resolution for point sources (\fig{astig}).  

The astigmatism correction is derived from observations of GCRV~12336, the central star of the Dumbbell Nebula (M~27), which exhibits \htwo\ absorption features across the \fuse\/ waveband. We cross-correlate and combine the absorption features from a small range in X, fit a parabola to each set of combined features, compute the shift required to straighten each parabola, and interpolate the shifts across the waveband.  Because an astigmatism correction has been derived only for point sources, no correction is performed on the spectra of extended sources, airglow spectra, or observations with APERTURE = RFPT.  

The correction is stored in the calibration file ASTG\_CAL as a two-dimensional image representing the region of the detector containing the target spectrum.  A separate image is provided for each aperture.  The value of each image pixel is the astigmatism correction ($\Delta X$ in pixels) to be applied to that pixel.  The entire image is shifted in Y to match the centroid of the target spectrum, and the appropriate correction is applied to each photon event in the target aperture.  The corrected X coordinates are not written to the IDF, but are passed immediately to the wavelength-assignment subroutine.  In effect, we apply a two-dimensional wavelength calibration, which depends upon both the X and Y coordinates of each photon event.

\subsubsection{Assign Wavelengths}

The wavelength calibration is stored as a binary table extension in the calibration file WAVE\_CAL (\S\ \ref{sec_wavecal}); a separate extension is provided for each aperture.  Wavelengths are tabulated for integral values of X, assumed to be in motion-corrected FARF coordinates.  Given the astigmatism-corrected X and CHANNEL arrays, the subroutine {\tt cf\_dispersion} considers each aperture in turn and reads the corresponding calibration table.  It interpolates between tabulated values of X to derive the wavelength of each photon event.  Photon events not assigned to an aperture (CHANNEL = 0) are not wavelength calibrated.

\subsubsection{Doppler Correction}

The component of the spacecraft's orbital velocity in the direction of the target is stored in the ORBITAL\_VEL array of the timeline table.  The component of the earth's orbital velocity in the direction of the target is stored in the IDF file header keyword V\_HELIO.  Their sum is used to compute a time-dependent Doppler correction, which is applied to each photon event by the subroutine {\tt cf\_doppler\_and\_heliocentric}.  The resulting wavelength scale is heliocentric.  Because histogram data are assigned identical arrival times, their Doppler correction is not time dependent.  Histogram exposures are kept short (approximately 500 seconds) to minimize the resulting loss of spectral resolution.  The final wavelength assigned to each photon event is stored in the LAMBDA array of the IDF.

\subsection{Flux Calibration}\label{flux}

Because the instrument sensitivity varies through the mission, we employ a set of time-dependent effective-area files (AEFF\_CAL).  We interpolate between the two files whose dates bracket the exposure start time but do not extrapolate beyond the most recent effective-area file.  Within each calibration file, the instrumental effective area is stored as a binary table extension, with a separate extension provided for each aperture. The module {\tt cf\_flux\_calibrate} invokes a single subroutine, {\tt cf\_convert\_to\_ergs}.  Considering each aperture in turn, the program reads the appropriate calibration files, interpolates between them if appropriate, and computes the ``flux density'' of each photon (in units of erg cm$^{-2}$) according to the formula
\begin{equation}
{\rm ERGCM2} =  {\rm WEIGHT} \times  h c \;  / \;  {\rm LAMBDA} \;  / \; A_{\rm eff}(\lambda),
\end{equation}
where ERGCM2, WEIGHT, and LAMBDA are read from the photon-event list (\S\ \ref{idf_format}), $h$ is Planck's constant, $c$ the speed of light, and $A_{\rm eff}(\lambda)$ the effective area at the wavelength of interest.  Only photon events assigned to an aperture are flux calibrated; events with CHANNEL = 0 are ignored.  The flux density computed for each photon event is stored in the ERGCM2 array of the IDF.  This array is not used by the pipeline, but is employed by some of our interactive IDF manipulation tools.

\subsection{Create Bad-Pixel Map}\label{sec_bpm}

When possible, spectra are extracted using an optimal-extraction algorithm (\S\ \ref{sec_optimal}) that employs a bad-pixel map (BPM) to correct for flux lost to dead spots and other detector blemishes.  Because motions of the spacecraft and its optical components cause \fuse\/ spectra to wander on the detector, a particular spectral feature may be affected by a dead spot for only a fraction of an exposure.  We thus generate a bad-pixel map for each exposure.  The module {\tt cf\_bad\_pixels} reads a list of dead spots from a calibration file (QUAL\_CAL), determines which of them overlap the target aperture, tracks the motion corrections applied to each, and converts the motion-corrected coordinates to wavelengths.  The resulting bad-pixel map (identified by the suffix ``bpm.fit'') has a format similar to that of an IDF (\S\ \ref{sec_bpm_format}), but the WEIGHT column, whose values range from 0 to 1, represents the fraction of the exposure that each pixel was affected by a dead spot.  No BPM file is created if the (screened) exposure time is less than 1 second.  BPM files are not archived, but can be generated from an IDF and its associated jitter file using software distributed with CalFUSE (available from MAST).  

\subsection{Extract Spectra}\label{extract_spectra}

Through all previous steps of the pipeline, we resist the temptation to convert the photon-event list into an image.  In the module {\tt cf\_extract\_spectra}, we relent.  Indeed, we generate four sets of images: a background model, a bad-pixel mask, a two-dimensional probability distribution of the target flux, and a spectral image for each extracted spectrum (LiF and SiC).  Only photon events that pass all of the requested screening steps (\S\ \ref{sec_screen}) are considered.  If the (screened) exposure time is less than 1 second or no photon events survive the screening routines, then the program generates a null-valued spectral file.

\subsubsection{Background Model}\label{sec_background}

\begin{figure}
\epsscale{.80}
\plotone{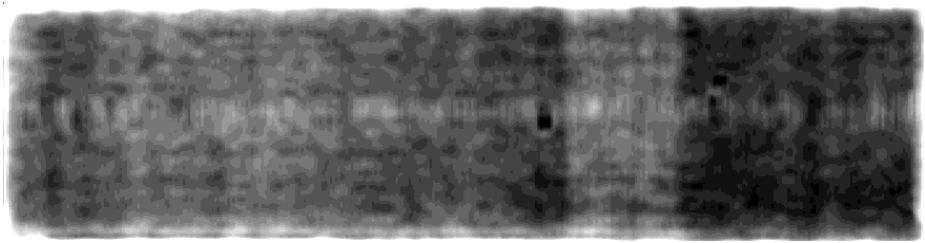}
\caption{Night-time scattered-light image for detector 1A.  Note the vertical scattered-light stripe to the right of the image center.}
\label{fig_bkgd}
\end{figure}

Microchannel plates contribute to the detector background via beta decay of $^{40}$K in the MCP glass. On orbit, cosmic rays add to this intrinsic background to yield a total rate of $\sim$ 0.5 counts cm$^{-2}$ s$^{-1}$.  Scattered light, primarily geocoronal Lyman $\alpha$, contributes a well-defined illumination pattern (\fig{fig_bkgd}) that varies in intensity during the orbit, with detector-averaged count rates as small as 20\% of the intrinsic background during the night and 1-3 times the intrinsic rate during the day.  We assume that the observed background consists of three independent components, a spatially uniform dark count and spatially-varying day- and nighttime scattered-light components.  Properly scaling them to the data is thus a problem with three unknown parameters.  We attempt to fit as many of these parameters as possible directly from the data.  When such a fit is not possible, we estimate one or more components and fit the remainder.

Background events due to the detector generally have pulse-height values lower than those of real photons (\S\ \ref{sec_pha}).  The observed dark count thus depends on the pulse-height limits imposed on the data.  An initial estimate of the dark count, as a function of the lower pulse-height threshold, is read from the background characterization file (BCHR\_CAL).  The day and night components of the scattered-light model are read from separate extensions of the appropriate time-dependent background calibration file (BKGD\_CAL).  The background models are scaled to match the counts observed on unilluminated regions of the detector.  The Y limits of these regions (selected according to the target aperture) are read from header keywords in the IDF.  Airglow photons in these regions are excluded from the analysis.  The day and night counts in the background regions of the detector are summed and recorded.

In its default mode, the subroutine {\tt cf\_scale\_bkgd} estimates the uniform background as follows:  The background regions of the day and night scattered-light models are scaled by their relative exposure times, summed, and projected onto the X axis (to produce a histogram in X).  A similar histogram (called the ``empirical background spectrum'') is constructed from the data.  An iterative process is used to determine the uniform component and scattered-light scale factor that best reproduce the observed X distribution of background counts.  The uniform component is then subtracted from the day and night totals computed above, and the day and night components of the scattered-light model are scaled to match the remaining observed counts. 

If the empirical background spectrum is too faint, we do not attempt to fit it, but assume that the uniform component of the background is equal to the tabulated dark-count rate scaled by the exposure time.  The day and night components of the scattered-light model are calculated as above.  Users who require a more accurate background model may wish to combine data from multiple exposures before extracting a spectrum (see \S\ \ref{sec_multiple}).
If the empirical background spectrum is very bright -- as, for example, when nebular emission or a background star contaminates one of the other apertures (and thus the background-sample region) -- no fit is performed.  Instead, the day and night components of the scattered-light model are scaled by the day and night exposure times, the tabulated dark-count rate is scaled by the total exposure time, and the three components are summed to produce a background image. This scheme is also used for histogram data.  Because histogram files contain data only from the region about the desired extraction window, the background cannot be estimated from other regions of the detector.  Fortunately, histogram observations typically consist of short exposures of bright targets, so the background is comparatively faint.

Our day and night scattered-light models were derived from the sum of many deep background observations spanning hundreds of kiloseconds.  Individual exposures that differ markedly from the mean were excluded from the sum.  The data were processed only through the FARF-conversion and data-screening steps of the pipeline.  Airglow features were replaced with a mean background interpolated along the dispersion (X) axis of the detector.  An estimate of the uniform background component was subtracted from the final image, and the data were binned by 16 detector pixels in X.  This process was performed on both day- and night-only data sets.  We produce a new set of background images every 6 to 12 months, as the effects of gain sag and adjustments of the detector high voltage slowly alter the relative sensitivity of the illuminated and background regions of the detectors.
 
{\it Caveats:}  While early versions of the pipeline (through v2.4) assume a 10\% uncertainty in the background flux, propagating it through to the final extracted spectrum, CalFUSE v3 treats uncertainties in the background as systematic errors and does not include them in the (purely statistical) error bars of the extracted spectra.

The algorithm assumes that the intensity of the uniform background is constant throughout an exposure.  This would be the case if it were due only to the detector dark count, but in fact the uniform background includes a substantial contribution from the scattered light and is thus brighter during day-time portions of an exposure.  The assumption of a constant uniform background can lead the algorithm to over-estimate the scale factors for both the uniform and spatially-varying components of the background model.  A better scheme would be to fit the day and night components of the uniform background separately.  
Similarly, when the empirical background spectrum is very faint, adopting the tabulated value of the uniform background is not the best solution.  It is likely that the scattered-light component of the uniform background would be better estimated from the observed day and night backgrounds.  The difference between the tabulated and observed levels of the uniform background will become greater as the mission extends into solar minimum and the intensities of both individual airglow features and the scattered-light component of the uniform background continue to weaken.

Grating scattering of point-source photons along the dispersion direction is potentially significant and is not corrected by the CalFUSE pipeline.  Typical values are 1--1.5\% of the continuum flux in the SiC channels and 10 times less in the LiF channels.  

\subsubsection{Probability Array}\label{sec_probability}

The optimal-extraction algorithm (\S\ \ref{sec_optimal}) requires as input a two-dimensional probability array representing the distribution of flux on the detector.  Separate probability arrays, derived from high signal-to-noise stellar observations, have been computed for each channel and stored as image extensions in the weights calibration file (WGTS\_CAL).  

By construction, the Y dimension of the probability array represents the maximum extent of the extraction window for a particular aperture.  For simplicity, all arrays employed by the optimal-extraction algorithm are trimmed to match the probability array in Y.  The centroids of the probability distribution and the target spectrum (recorded in the corresponding file headers) are used to determine the offset between detector and probability coordinates.  In the X dimension, all arrays are binned to the output wavelength scale requested by the user.  Default wavelength parameters for each aperture are specified in the header of the wavelength calibration file (WAVE\_CAL); the default binning for all channels is 0.013 \AA\ per output spectral bin, which corresponds to approximately two detector pixels.  The background array, originally binned by 16 pixels in X, is rescaled to the width of each output wavelength bin by the subroutine {\tt cf\_rebin\_background}.  The probability array is rescaled by the subroutine {\tt cf\_rebin\_probability\_array} to have a sum of unity in the Y dimension for each wavelength bin.

\subsubsection{Bad-Pixel Mask}\label{sec_make_mask}

A bad-pixel mask with the same wavelength scale and Y dimensions as the probability array is constructed from the BPM file (\S\ \ref{sec_bpm}) by the subroutine {\tt cf\_make\_mask}.  The array is initialized to zero.  For each entry in the BPM file, the value of the WEIGHT array is added to the corresponding pixel of the bad-pixel mask.  The mask is then normalized and inverted, so that the center of the deepest dead spot has a value of 0 and the regions outside are set to 1.  The conversion from pixel to wavelength coordinates can open gaps in the mask, which appear as values of unity surrounded by pixels with lower values.  We search for array elements that are larger than their neighbors and replace them with the mean value of the adjoining pixels.  If the BPM file is absent or a particular aperture is free of bad pixels, all elements of the bad-pixel mask are set to 1. 

\subsubsection{Optimal (Weighted) Spectral Extraction}\label{sec_optimal}

The extraction subroutine, {\tt cf\_optimal\_extraction}, is called separately for each of the two target spectra (LiF and SiC).  Inputs include the photon-event list and the indices of events that pass through the various screenings, as well as the 2-D background, probability, and bad-pixel arrays described above.  For numerical simplicity, extraction is performed using the WEIGHT of each photon event, rather than its ERGCM2 value.  A pair of 2-D data and variance arrays with the same dimensions as the probability array are constructed from the good photons whose CHANNEL values correspond to the target aperture.  For time-tag data, this process is straightforward: the LAMBDA and Y values of each photon event correspond to a particular cell in the data and variance arrays.  That cell in the data array is incremented by the photon weight, while the corresponding cell in the variance array is incremented by the square of the weight.  A 1-D raw-counts spectrum (useful for the statistical analysis of low count-rate data) is constructed simultaneously: for each photon event added to the data array, the appropriate bin of the counts spectrum is incremented by one.

For histogram data, the process is more complex, because the original detector image is generally binned by 8 pixels in Y and because each entry in the photon-event list represents the sum of many individual photons.  In the Y dimension, an event's WEIGHT is divided among 8 pixels (or the actual Y binning for that exposure, if different) according to the distribution predicted by the probability array.  In the X dimension, each event is assumed to have a width in wavelength space equal to the mean dispersion per pixel for the channel (read from the DISPAPIX keyword of the WAVE\_CAL file), and the WEIGHT of an event that spans the boundary between two output wavelength bins is divided between them.  This smoothing in X helps to mitigate the ``beating'' that would otherwise occur between detector pixels and output wavelength bins.

One-dimensional background, weights, and variance spectra are then extracted from the two-dimensional background, data, and variance arrays.  To insure that the three spectra sample the same region of the detector, only cells in the 2-D arrays for which the corresponding cell in the probability array has a value greater than $10^{-4}$ are included in the sum.  These limits differ slightly from those defined in the aperture (CHID\_CAL) calibration files.  As a result, the ratio of the final weights and counts spectra may not be a constant.  (Ideally, their ratio would equal the mean dead-time correction for the exposure.)  An initial flux spectrum, equal to the difference of the weights and background spectra, is used as input to the optimal-extraction algorithm.

CalFUSE employs the optimal-extraction algorithm described by \citet{Horne:86}, which requires as input the 2-D data, background, probability, and bad-pixel arrays and the 1-D initial flux spectrum.  Originally designed for CCD spectroscopy, the algorithm has been modified for the \fuse\/ detectors.  Specifically, instead of constructing a 2-D spatial profile from each data set, we use a tabulated probability array; the 2-D cosmic-ray mask, an integer array in the original algorithm, is replaced with the bad-pixel mask, which is a floating-point array; and the 2-D variance estimate is scaled by the bad-pixel mask.  Extraction is iterative: in the original version, iteration is performed until the cosmic-ray mask stops changing.  In our version, iteration continues until the output flux spectrum changes by less than 0.01 counts in all pixels.  If the loop repeats 50 times, the algorithm fails.  The number of iterations performed is written to the OPT\_EXTR keyword of the output file header.

If optimal extraction is successful, the variance of the optimal spectrum is computed using the recipe of \citet{Horne:86}.  We have adapted this recipe to produce weights and background spectra such that FLUX = WEIGHTS $-$ BKGD.  The resulting background spectrum is not smooth.  Optimal extraction is not performed on the spectra of extended sources or on those for which the quality of the computed spectral centroid is not HIGH.  (Both the centroid and its quality flag are stored in file header keywords.)  In these cases, or if the optimal-extraction algorithm fails, the initial flux, variance, weights, and background spectra are adopted.

However they are constructed, the final FLUX, ERROR (equal to the square root of the variance), WEIGHTS, and BKGD arrays (all in units of counts) are returned to the calling routine.  For time-tag data, the COUNTS array as described above is returned.
For histogram data, the COUNTS array is computed by dividing the final WEIGHTS array by the mean dead-time correction, which is stored in a file-header keyword TOT\_DEAD.
Also returned is the QUALITY array.  It is the product of the probability array and the bad-pixel map, projected onto the wavelength axis and expressed as an integer between 0 and 100.  Its value is 0 if all the flux in a wavelength bin is lost to a detector dead spot, 100 if no flux is lost. 

{\it Caveats:} The optimal-extraction algorithm is designed to improve the signal-to-noise ratio of the spectra of faint point sources.  Unfortunately, it is in precisely these cases that the spectral centroid is most likely to be uncertain.  Because proper positioning of the probability array is essential to the weighting scheme, observers of faint targets may wish to combine the IDFs from multiple exposures and re-compute their spectral centroids before attempting optimal extraction. 

\subsubsection{Extracted Spectral Files}\label{sec_fcal}

Because the optimal-extraction routine returns the final FLUX and ERROR arrays in units of counts, the spectral-extraction module applies a flux calibration to both arrays using the subroutine {\tt cf\_convert\_to\_ergs} described in \S\ \ref{flux}.  Dividing each array element by (EXPTIME $\times$ WPC), where EXPTIME is the length in seconds of the (screened) exposure and WPC the width in \AA ngstroms of each output wavelength bin, completes the conversion to units of \flux.  The format of the extracted spectral files is described in \S\ \ref{spec_format}.

\subsection{Trailer and Image Files}\label{sec_trailer}

A number of supplementary files are generated by CalFUSE and archived with the data.  For each exposure and detector segment, the pipeline generates a trailer file and a pair of image files in Graphics Interchange Format (GIF).  The trailer file (suffix ``.trl'') contains timing information for all pipeline modules and any warning or error messages that they may have generated.  The first image file contains an image of the detector overlaid by a wavelength scale and extraction windows for each aperture (suffix ``ext.gif'').   Only photon events flagged as good are included in the plot, unless there are none, in which case all events are plotted.  The second image file presents count-rate plots for both the LiF and SiC target apertures (suffix ``rat.gif'').  These arrays come from the timeline table in the IDF and exclude photons flagged as airglow.  These image files are powerful tools for diagnosing problems in the data, revealing, for example, when high background levels cause the SiC1 LWRS extraction window to be misplaced.

\subsection{Observation-Level Files}\label{obs_files}

For each exposure, CalFUSE produces LiF and SiC spectra from each of four detector segments, for a total of eight extracted spectral files.  OPUS combines them into a set of three observation-level files for submission to MAST.  Observation-level files are distinguished from exposure-level files by having an exposure number of 000.  Depending on the target and the scientific questions at hand, these files may be of sufficient fidelity for scientific investigation. Here is a brief description of their contents:

ALL: For each combination of detector segment and channel (LiF1A, SiC1A, \etc), we combine data from all exposures in the observation into a single spectrum. If the individual spectra are bright enough, we cross correlate and shift them before combining.  
(For each channel, the shift calculated for the detector segment spanning 1000--1100 \AA\ is applied to the other segment as well.)  If the spectra are too faint for cross correlation, we combine the individual IDFs and extract a single spectrum to optimize the background model.  Combined spectra (WAVE, FLUX, and ERROR arrays) for each of the eight channels are stored in separate binary table extensions in the following order: 1ALIF, 1BLIF, 2BLIF, 2ALIF, 1ASIC, 1BSIC, 2BSIC, and 2ASIC.

ANO (all, night-only): With the same format as the ALL files, these spectra are constructed using only data obtained during the night-time portion of each exposure. They are generated only for time-tag data, and only if EXPNIGHT $> 0$.  The shifts calculated for the ALL files are applied to the night-only data; they are not recomputed.

NVO (National Virtual Observatory): These files contain a single spectrum spanning the entire \fuse\/ wavelength range. The spectrum is assembled by cutting and pasting segments from the most sensitive channel at each wavelength. Segments are shifted to match the guide channel (either LiF1 or LiF2) between 1045 and 1070 \AA. Columns are WAVE, FLUX, and ERROR and are stored in a single binary table extension.

The ALL file is used to generate a ``quick-look'' spectral plot for each observation.  When available, combined spectra from channels spanning the \fuse\/ waveband are plotted in a single GIF image file (suffix ``specttagf.gif'' or ``spechistf.gif'').  This plot appears on the MAST preview page of each observation.  Four additional GIF files contain the combined LiF1, LiF2, SiC1, and SiC2 spectra for each observation.

{\it Caveats:}  Cross-correlation may fail, even for the spectra of bright stars, if they lack strong spectral features.  Examples are nearby white dwarfs with weak interstellar absorption lines.  If cross-correlation fails for a given exposure, that exposure is excluded from the sum.  Thus, the exposure time for a particular segment in an ALL file may be less than the total exposure time for that observation.

The cataloging software used by MAST requires the presence of an ALL file for each exposure, not just for the entire observation. We generate exposure-level ALL files, but they contain no data, only a FITS file header. The observation-level ALL files discussed above can be distinguished by the string ``00000all'' in their names.

\subsection{Quality Control and Archiving}

Before the reduced data are archived, they undergo a two-step quality-control process:  
First, a set of automated checks is performed on each exposure.  
The software compares the flux observed in the guide channel (LiF1 or LiF2) with that expected for the target and with that observed in the other three channels.
If an anomaly is detected, a flag is set requesting manual investigation.  
The software works well for bright continuum sources, but often flags faint or emission-line targets as 
unsuccessful observations. 
Second, a member of the \fuse\/ operations team investigates any warnings generated by the software.
If it is determined that less than 50\% of the requested data were obtained, the target is re-observed.  

The philosophy of the \fuse\/ project is to archive data whenever possible, even if it does not satisfy the requirements of the original investigator.  As a result, the MAST archive contains a number of \fuse\/ data sets that are in some way flawed (\eg, misaligned channels, partial loss of guiding, or no good observing time).  Users should be aware of this possibility.

If the pipeline detects an error in the data or its associated housekeeping or jitter files, it writes a warning to both the trailer file and the headers of the IDF and extracted spectral files.  Users are advised to scan trailer files for the ``WARNING'' string and spectral files for ``COMMENT'' records.  Occasionally, the \fuse\/ operations team inserts comments directly into the headers of raw data files.  Such comments may warn of an unusual instrument configuration, errors in the reported target coordinates, or data obtained during slews.

Observation names beginning with the letter ``S'' are science-verification observations and may have been obtained with an unusual instrument configuration.  For example, the program S523 was designed to test procedures for observing a bright object by defocusing the SiC mirrors.  The LiF mirrors were moved for some S523 exposures.  As a result, data from this program should be used with caution.  Abstracts for all \fuse\/ observing programs are available from MAST.
 
\section{CALIBRATION FILES}\label{sec_calfiles}

\subsection{Wavelength Calibration}\label{sec_wavecal}

\subsubsection{Derivation of the \fuse\/ Wavelength Calibration}

Our principal wavelength-calibration target is GCRV~12336, central star of the planetary nebula M~27, whose spectrum exhibits a myriad of molecular-hydrogen absorption features \citep{McCandliss:07}.  For the SiC1B channel, these data are supplemented at the shortest wavelengths by spectra of the hot white dwarf G~191-B2B \citep{Lemoine:02}.  Spectra obtained through each of the three \fuse\/ apertures were fully reduced, corrected for astigmatism, and used to derive an empirical mapping of pixel to wavelength.  For each channel, standard optical expressions were used to derive a theoretical dispersion solution, which was fit to the empirical data with only its constant term (the zero-point of the wavelength scale) as a free parameter.  The shifted theoretical dispersion solution was used to generate the wavelength-calibration file.

\begin{figure}
\epsscale{.80}
\plotone{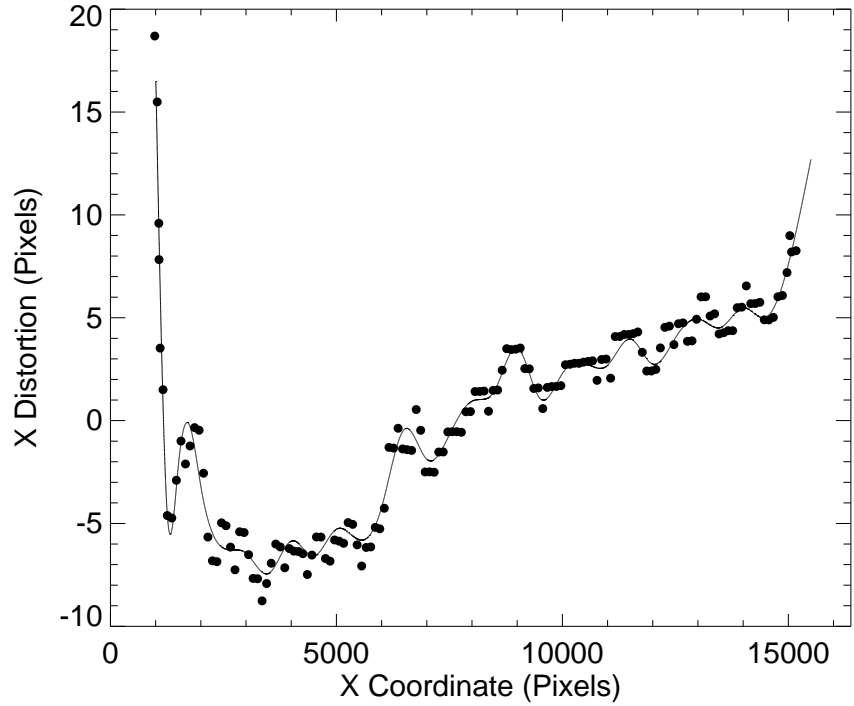}
\caption{Geometric distortion in the X coordinate of the LiF1A LWRS channel.  Data represent the difference between the measured locations of \htwo\ lines in the spectrum of GCRV~12336 and those predicted by a theoretical dispersion relation.  The solid line is a spline fit to the residuals.}
\label{fig_residuals}
\end{figure}

Early versions of the pipeline relied on the wavelength calibration to correct for non-linearities in the detector X scale (the geometric distortion discussed in \S\ \ref{sec_geometric}).  Correcting for this effect separately has greatly improved the accuracy of the \fuse\/ wavelength scale.
To determine the geometric distortion in the X dimension, a spline was fit to the residuals from each aperture (expected minus observed X coordinate of each absorption feature; \fig{fig_residuals}).  In practice, residuals from all six apertures (both LiF and SiC channels) were included in the fit, but data from the other five apertures were weighted 100 times less than those for the aperture being fitted. The additional data points help to constrain the fit in wavelength regions where the data are sparse or missing.  The spline fits from all six apertures were then used to construct the two-dimensional map of detector distortions in the X dimension that is used by the geometric-distortion routine described in \S\ \ref{sec_geometric}.  The process is iterative, with the residuals (ideally) becoming smaller with each iteration.

The scatter of individual measurements about the spline fit is caused in some cases by blended absorption lines and in others by localized distortions induced by the fiber-bundle structure of the MCPs. This scatter is thus a fair estimate of the inaccuracies that the user may expect in the relative measurement of the wavelength of any given feature.  The wavelength inaccuracies caused by localized distortions are 3--4 detector pixels (0.025 \AA\ or 7 \kms) at most wavelengths, but may be as large as 6--8 pixels. They occur in tiny windows about 1 to 3 \AA\ wide, depending on the channel and segment. Some data sets show larger residuals. These distortions are inherent in the \fuse\/ data set and represent the ultimate limit to the accuracy of the \fuse\/ wavelength calibration.

\subsubsection{Zero-Point Uncertainties}

The \fuse\/ wavelength calibration assumes that the motion-corrected spectrum falls at a precise location on the detector. If it is shifted in X, then the wavelength scale of the extracted spectrum will suffer a zero-point offset.  For the guide channel (either LiF1 or LiF2), the dominant source of wavelength errors is thermally-induced rotations of the spectrograph gratings, which depend on the satellite attitude.  For the other channels, additional wavelength errors come from mirror misalignments that shift the target away from the center of the aperture.  Such misalignments may produce zero-point offsets of up to $\pm 0.15$ \AA\ for point sources in the LWRS aperture.  Offsets are less than $\pm 0.02$ \AA\ for the MDRS aperture and are negligible for the HIRS aperture.

We define the zero point of our wavelength scale by requiring that the Lyman $\beta$ airglow feature, observed through the HIRS aperture and processed as if it were a point source, be at rest in spacecraft coordinates when all Doppler corrections are turned off.  The use of an airglow feature eliminates errors due to mirror motions in the non-guide channels, but not errors in the grating-motion correction, so we measure the Lyman $\beta$ line in some 200 background exposures and shift their mean velocity to 0 \kms.  For each channel, all three apertures (HIRS, MDRS, and LWRS) use this HIRS-derived wavelength scale (WAVE\_CAL version 022 and greater).

The grating-motion correction (\S\ \ref{sec_grating_motion}) is designed to place the centroid of each Lyman $\beta$ airglow feature at a fixed location in FARF coordinates.  On average, it achieves that goal: for our sample of 200 background exposures, the measured velocity of the Lyman $\beta$ line has a standard deviation of between 2 and 3 \kms, depending on the channel.  Unfortunately, some combinations of pole and beta angle are not well corrected, leading to velocity offsets of 10 \kms\ or more, and additional motions -- of either the gratings or some other optical component -- can shift the extracted spectra by several \kms\ from one exposure to the next.

\begin{figure}
\epsscale{.80}
\plotone{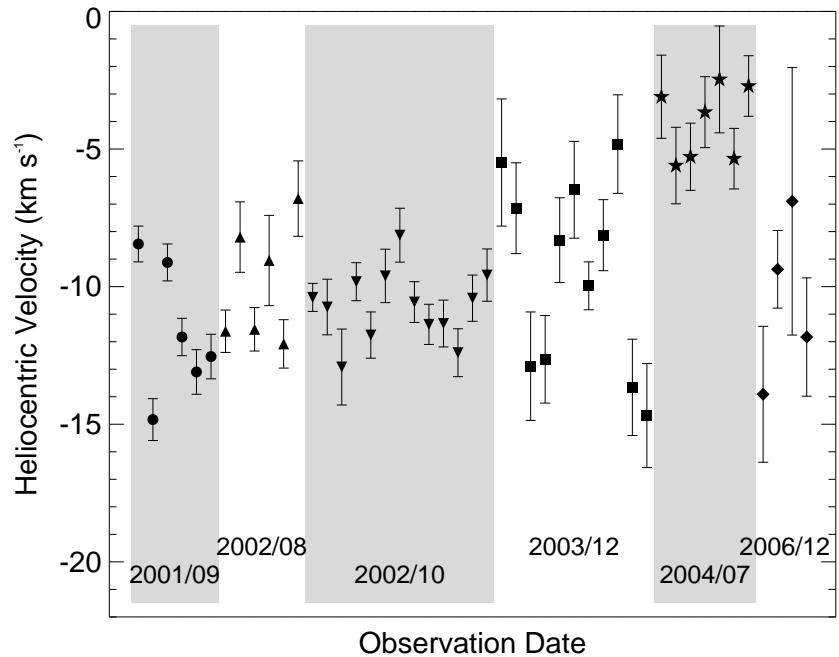}
\caption{Measured heliocentric velocity of the interstellar \oone\ $\lambda 1039.23$ absorption feature in each of 47 exposures of the white dwarf KPD 0005+5106 through the high-resolution (HIRS) aperture.  \citealt{Holberg:98} report a heliocentric velocity of $-7.50 \pm 0.76$ \kms\ for the interstellar features along this line of sight.}
\label{fig_velocity}
\end{figure}

Figure \ref{fig_velocity} presents the measured wavelength of the interstellar \oone\ $\lambda 1039$ absorption feature in 47 exposures of the hot white dwarf KPD 0005+5106 obtained through the HIRS aperture.  The 2001 and 2002 data show little scatter and yield a mean velocity of $-10.7 \pm 1.9$ \kms.  (\citealt*{Holberg:98} report a heliocentric velocity of $-7.50 \pm 0.76$ \kms\ for the interstellar features along this line of sight.)  The 2003 data (all from a single observation) span nearly 10 \kms.  The 2004 data (again from a single observation) are tightly correlated but offset by $\sim$ 7 \kms.  These data were obtained at a spacecraft orientation (beta and pole angles) that is generally well corrected by our grating-motion algorithm; apparently, some other effect is at work.  The 2006 data come from the LiF2 channel, which became the default guide channel in 2005 (\S\ \ref{sec_fesb}).

We do not recommend the general use of airglow lines to fix the absolute wavelength scale of point-source spectra for several reasons: First, airglow emission fills the aperture, so the resulting airglow lines provide no information about the position of the target relative to the aperture center. Second, the jitter correction (for all channels) and the mirror-motion correction (for the SiC channels) are inappropriate for airglow emission.  Third, the Doppler correction for the spacecraft's orbital motion can degrade their resolution.

\subsubsection{Diffuse Emission}

The \fuse\/ wavelength scale is derived from astigmatism-corrected, point-source spectra. Extended-source (diffuse) spectra are not corrected for astigmatism. If point-source data are processed with the astigmatism correction turned off, the resulting wavelength errors are less than about 4 detector pixels, consistent with the uncertainties in the wavelength scale. Therefore, the present \fuse\/ wavelength calibration should be adequate for extended-source spectra.  Airglow lines are useful for determining the zero-point for extended sources that fill the aperture.

%\clearpage

\begin{deluxetable}{lccc}
\tablecolumns{4}
\tablewidth{0pt}
\tablecaption{Stellar Parameters Adopted for \fuse\/ Flux Standards\label{tab_fluxcal}}
\tablehead{
& \colhead{$T_{\rm eff}$} & \colhead{$\log g$} & \colhead{$V_{\rm rad}$} \\
\colhead{Name} & \colhead{(K)} & \colhead{(cm s$^{-2}$)} & \colhead{(\kms)}
}
\startdata
GD~71 &  32,843 &  7.783 &  \phs80.0 \\
GD~659 & 35,326 &  7.923 &  \phs33.0 \\
GD~153 & 39,158 &  7.770 &  \phs50.0 \\
HZ~43  & 50,515 &  7.964 &  \phs20.6 \\
GD~246 & 53,000 &  7.865 & $-13.2$ \\
G~191-B2B  & 61,200 & 7.5 &  \nodata
\enddata
\tablecomments{For G~191-B2B, we use the model employed by \citealt{HUT2CAL2} for the final {\it Astro-2} calibration of HUT.}
\end{deluxetable}

\subsection{Flux Calibration}\label{sec_fluxcal}

\subsubsection{Derivation of the Effective-Area Curve}

The \fuse\/ flux calibration is based on in-flight observations of
the well-studied DA (pure-hydrogen) white dwarfs listed in Table
\ref{tab_fluxcal}, which have been observed at regular intervals
throughout the mission.  For each channel, data from multiple stars
are combined to track changes in the instrument sensitivity using
a technique similar to that developed by \citet{Massa:00} for the
{\it International Ultraviolet Explorer (IUE)}/ satellite.  The algorithm yields a series
of time- and wavelength-dependent sensitivity curves as well as the
spectrum of each star, in units of raw counts, as it would have
appeared on a date early in the mission, which we choose
to be $T_0 = 1999$ December 31.  
(We refer to the latter as ``$T_0$ spectra.'')  

For each star, we generated a synthetic spectrum using the programs
TLUSTY (version 200) and SYNSPSEC (version 48) of \citet{Hubeny:Lanz:95}. 
The non-LTE pure-hydrogen model atmospheres were computed according to a prescription by Hubeny (private communication) using 200 atmospheric layers to ensure an optimal absolute flux accuracy. The atmospheric parameters listed in Table \ref{tab_fluxcal},
consistent with {\it HST, IUE,} and optical observations, were
used to compute the models (Holberg, private communication). 
For G~191-B2B, we
used the model employed by \citet{HUT2CAL2} for the final {\it Astro-2}
calibration of HUT.  
Observations of these stars with the Faint Object Spectrograph aboard \hst\/
have shown that the models, including parameter uncertainties, are
consistent to within 2\% at wavelengths longer than Lyman $\alpha$
\citep*{Bohlin:95, Bohlin:96}. Uncertainties in the far-ultraviolet
waveband are slightly higher, as discussed by \citet{HUT2CAL2}.

\begin{figure}
%\epsscale{.80}
\plotone{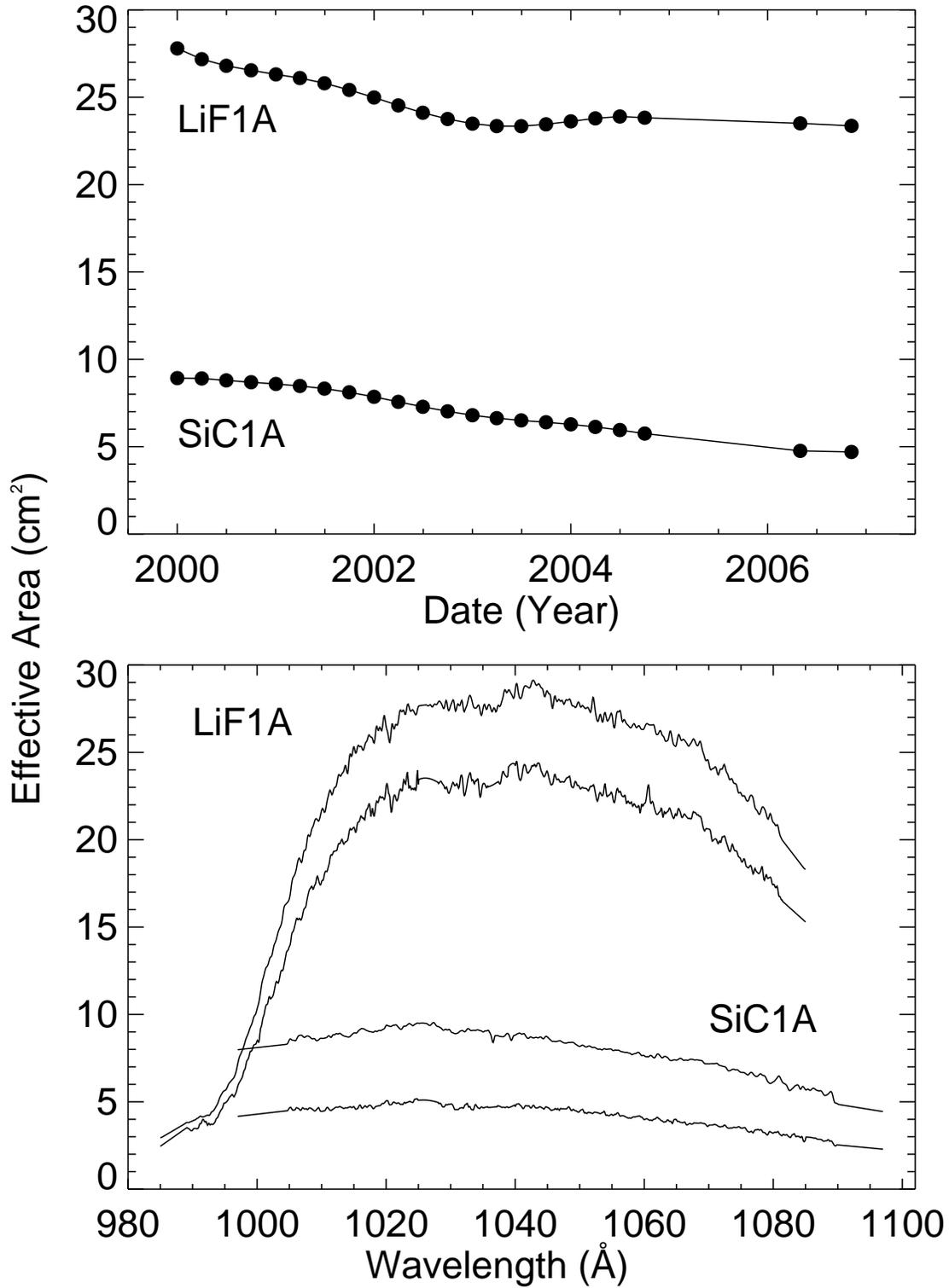}
\caption{\fuse\/ sensitivity as a function of time.  {\em Upper panel:}\/ Effective area of the LiF1A and SiC1A channels, averaged over the wavelength region 1030--1040 \AA.  The gap between 2004 October and 2006 May represents the period after the loss of the third reaction wheel, when few calibration targets were observed.  {\em Lower panel:}\/  Effective-area curves for the LiF1A and SiC1A channels, dated 1999 and 2006.  (For both channels, the 1999 curve has the higher effective area.)}
\label{fig_fluxcal}
\end{figure}

For each channel, the effective area in units of cm$^2$ is computed
by dividing one or more $T_0$ spectra in units of counts s$^{-1}$
\AA$^{-1}$ by a synthetic white-dwarf spectrum in units of photons
cm$^{-2}$ s$^{-1}$ \AA$^{-1}$.  
We find excellent agreement between the effective areas derived from the
different standard stars.
Sensitivity curves for the LiF1A and SiC1A channels are presented in \fig{fig_fluxcal}.
(Effective-area curves for all \fuse\/ channels are available from MAST.)
The sensitivity of the LiF1A channel decreased by $\sim$ 15\%
over the first three years of the mission, but appears to have stabilized; that of the
SiC1 channel has declined by $\sim$ 45\% since launch and is falling still (though slowly).
Effective-area curves (AEFF\_CAL) for each channel and detector
segment were generated at three-month intervals until the loss of the third reaction wheel 
in 2004 December; we plan to generate them at six-month intervals for the duration of the mission.

{\it Caveats:} We do not attempt to correct spectra obtained through the MDRS and HIRS apertures for changes in instrument sensitivity, but employ a single effective-area curve for each.
The low throughput of these apertures, combined with the likelihood that their spectra are non-photometric, makes tracking changes in their sensitivity both more difficult and less useful than for the LWRS aperture.

\subsubsection{Systematic Uncertainties}

The greatest uncertainties in a line or continuum flux derived from
a \fuse\/ spectrum are due to systematic effects.  An estimate of
the uncertainty in our flux calibration can be obtained by comparing
the effective-area curves derived from different white-dwarf stars.
Differences among the curves reflect errors in both the model
atmospheres and the stellar parameters upon which they are based.
In most channels, the scatter in the derived effective areas is
between 2 and 4\%.

The photometric accuracy of \fuse\/ spectra is subject to numerous effects that cannot be fully corrected by the CalFUSE pipeline.  A target centered in an aperture of the guide channel (LiF1 or LiF2) may not be centered in the corresponding apertures of the other three channels.  Since the loss of the first two reaction wheels in 2001, spacecraft drifts may move the target out of even the guide-channel aperture.  While the pipeline does attempt to flag times when the target is out of the aperture, the algorithm used is conservative in that it underestimates the time lost to pointing errors (\S\ \ref{sec_flag_jitter}).  The user is advised to consult the count-rate plots generated by the pipeline (suffix ``rat.gif''; \S\ \ref{sec_trailer}) and the LIF\_CNT\_RATE and SIC\_CNT\_RATE arrays of the IDF timeline table to determine the photometric quality of an exposure.  Using tools available from MAST or the user-defined good-time intervals discussed in \S\ \ref{sec_user_defined}, users can reject time periods when the count rate is low or re-scale the flux of low-count-rate exposures.

When a point-source target falls near the top or bottom edge of an aperture, vignetting in the spectrograph may attenuate the target flux in a wavelength-dependent way.  Astigmatism gives \fuse\/ spectra the shape of a bow tie (\fig{detector1a}).  If vignetting is important, then the spectrum will lie below the center of the aperture on one side of the bow tie and above it on the other.  Significant flux loss is possible in wavelength regions far from the center of the bow tie.

Other systematic uncertainties are imposed by various detector flat-field
effects; their relative importance depends upon one's scientific goals. 
For narrow emission lines, flux uncertainties are dominated by
the moir\'{e} pattern (high-frequency ripples
due to beating among the arrays of microchannel pores in the MCP
stack; \S\ \ref{sec_moire}), unless the observation was obtained using an FP split or
the equivalent was achieved via grating and mirror motions.  For
broad features, the moir\'{e} is not important, but larger-scale
flat-field features are.  These effects are discussed in 
\anchor{http://archive.stsci.edu/fuse/dhbook.html#DetectorEffects}{{\em The
FUSE Instrument and Data Handbook.}}  Finally, when fitting a spectral
energy distribution, the greatest uncertainty is caused by worms (\S\ \ref{sec_worm}),
which may depress the observed flux over tens of \AA ngstroms by 50\% or more.

\subsubsection{Extended Sources}

The \fuse\/ flux calibration is derived from point-source targets.
Because the distribution of flux in the cross-dispersion direction
differs for point and extended sources, it is possible that the
instrumental sensitivity may also differ; this question has not
been explored in detail.  Extended spectra are less
affected by worms (\S\ \ref{sec_worm}) than are point-source spectra. Moreover, because
the spectrum of a diffuse emitter is spread over a larger region
of the detector, it will suffer less from local flat-field effects.

\section{DISCUSSION}\label{sec_discussion}

\subsection{Spacecraft Guiding on the LiF2 Channel}\label{sec_fesb}

The switch from FES A to FES B as the default guide camera in 2005 July has two principal effects on the quality of \fuse\/ data.
First, tracking with FES A ensured that targets remained in the center of the LiF1 aperture, which is the most sensitive channel in the astrophysically-important 1000--1100 \AA\ waveband.  
Tracking with FES B will keep targets centered in the LiF2 aperture, increasing the likelihood of data loss in the LiF1 channel.
Second, in order to optimize the optical focus of FES B, the LiF2 
FPA was moved out of the focal plane of the LiF2 primary mirror.  
Observations of point sources with the LWRS aperture are unaffected, and 
the point-source spectral resolution of this channel is unchanged, but the   
throughput of the narrow LiF2 apertures is reduced.
The effective transmission of the apertures has not been characterized in detail, but is approximately 70\% for LIF2 MDRS and 15\% for LiF2 HIRS, versus 98\% and 60\% for their LiF1 counterparts.
The spectral resolution for diffuse sources is expected to be slightly lower in LiF2  than in LiF1.  

\subsection{Scattered Solar Emission}\label{sec_solar}

In addition to airglow lines, scattered solar emission features are present in the SiC channels when observing at high beta angles during the sunlit portion of the orbit. Emission from \cthree\ $\lambda 977.0$, Lyman $\beta$ $\lambda 1025.7$, and \osix\ $\lambda \lambda 1031.9, 1037.6$ has been positively identified. Emission from \nthree\ $\lambda 991.6$ and \ntwo\ $\lambda 1085.7$ may also be present.
It is believed that sunlight is scattered by reflective, silver-coated Teflon blankets lying above the SiC baffles. At low beta angles, scattered solar emission is less apparent, because the blankets are shaded by the SiC baffles and the open baffle doors and because the radiation strikes the blankets at a high angle of incidence. It is unknown at which beta angle, if any, the solar emission completely disappears. Because the LiF channels lie on the shadowed side of the spacecraft, solar emission lines are not seen in LiF spectra.
\cthree\ and \osix\ emission observed in the SiC channels during orbital day should always be compared with the emission observed either with the LiF channel or during the nighttime portion of an orbit.

Since the failure of the third reaction wheel in 2004 December, \fuse\/ mission controllers have experimented with the use of non-standard roll angles to improve spacecraft stability. These roll angles can place the spacecraft in a configuration that greatly increases the sunlight scattered into one of the SiC channels.  The scattered light, mostly Lyman continuum emission, appears as an increase in the background at wavelengths shorter than about 920 \AA; strong, resolved Lyman lines are present at longer wavelengths.  When present, it is generally seen in only one of the two SiC channels.  We have no way to model or subtract this emission.

\subsection{The Worm}\label{sec_worm}

\begin{figure}
\epsscale{0.8}
\plotone{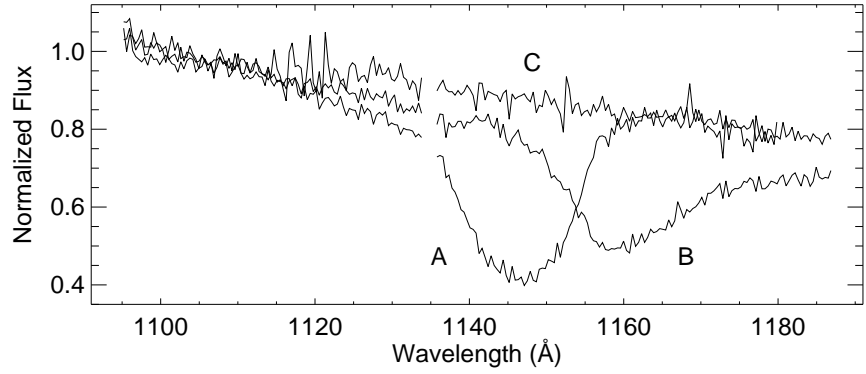}
\caption{Point-source spectra showing the effects of the worm.  Spectra A and B, obtained with the LiF1B channel, show deep depressions near 1145 and 1160 \AA, respectively.  The wavelength of maximum attenuation varies with the Y position of the target within the aperture.  Spectrum C, obtained with the LiF2A channel, is unattenuated.}
\label{fig_worm_spec}
\end{figure}

\begin{figure}
\epsscale{.80}
\plotone{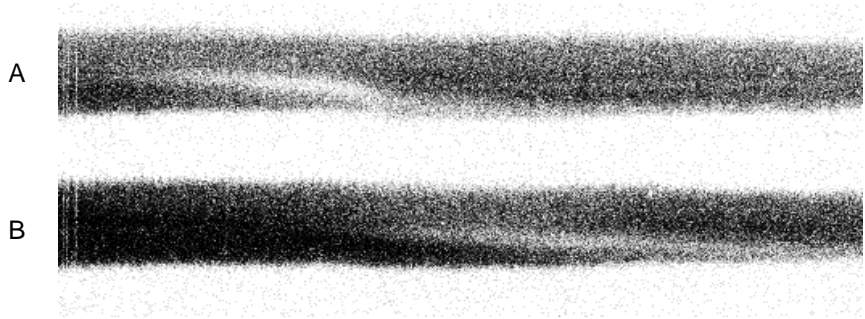}
\caption{Detector images showing the effects of the worm.  In these negative images, worms appear as bright stripes parallel to the dispersion axis.  The data shown correspond to spectra A and B in \fig{fig_worm_spec} and span wavelengths between 1134 and 1187 \AA.}
\label{fig_worm}
\end{figure}

The spectra of point-source targets occasionally exhibit a depression in flux that may span as much as 50 \AA\ (\fig{fig_worm_spec}).  These depressions appear in detector images as narrow stripes roughly parallel to the dispersion axis (\fig{fig_worm}).  The stripes, known as worms, can attenuate as much as 50\% of the incident light in affected portions of the spectrum.  Worms shift in the dispersion direction when the target moves in the aperture.  They are due to an unfortunate interaction between the horizontal focus of the spectrograph and the innermost wire grid (the quantum-efficiency grid; \S\ \ref{sec_geometric}).  Since the location of this focus point is a function of wavelength, the strength of a worm is exquisitely sensitive to the exact position of the spectrum on the detector.  We cannot determine this position with sufficient precision to correct reliably for flux lost to worms.  Though most prominent in LiF1B LWRS spectra, worms can appear in all channels and apertures.  Observers who require absolute spectrophotometry should carefully examine \fuse\/ spectral image files for the presence of worms.  The redundant wavelength coverage of the various \fuse\/ channels can be used to mitigate their effects.

\subsection{The Moir\'{e} Pattern in Histogram Data}\label{sec_moire}

\begin{figure}
\epsscale{.60}
\plotone{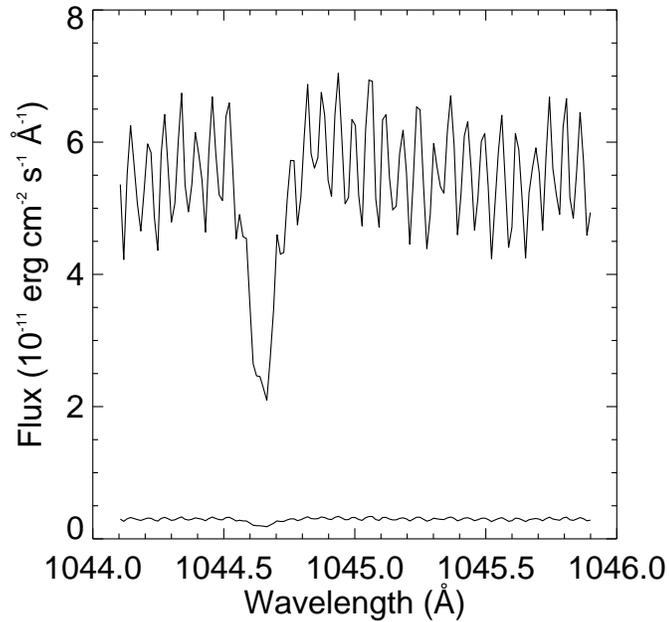}
\caption{Moir\'{e} pattern in the LiF2B spectrum of the star HD~209339, obtained in histogram mode.  The associated error array is overplotted.  The moir\'{e} ripples are strongest on this detector segment, but are also seen on segments 1A and 1B.}
\label{fig_noise}
\end{figure}

Since the release of CalFUSE v3.0, users have reported strong, non-Gaussian noise in the spectra of some bright stars observed in histogram mode.  An example is shown in \fig{fig_noise}.  The high-frequency ripples have a period of approximately 9 detector pixels, or about 0.06 \AA.  These ripples are a moir\'{e} pattern due to beating among the arrays of microchannel pores in the three layers of the MCP stack \citep{Tremsin:99}.  The moir\'{e} fringes are strongest on segment 2B, but are also visible on segments 1A and 2B.  The motion corrections applied to time-tag data tend to smooth out this effect, but it can be quite strong in histogram data.   Where it is present, users are advised to smooth or bin their spectra by at least one resolution element to reduce its effects.  This and other detector artifacts are described in the \anchor{http://archive.stsci.edu/fuse/dhbook.html#Moire}{{\em FUSE Instrument and Data Handbook}}.

\subsection{A Note about Time}

The \fuse\/ spacecraft uses Coordinated Universal Time (UTC).  
The spacecraft clock is updated periodically from the ground using a procedure that corrects for the signal transit time from the ground station to the spacecraft.  The ground station time comes from GPS satellites.
The Instrument Data System receives a 1 Hz signal from the spacecraft that is used to align the IDS clock with the spacecraft clock to an accuracy of $\pm 5$ ms.
In time-tag mode, the IDS typically inserts a time stamp into the data stream once per second, but can insert time stamps as frequently as 125 times per second.  Unfortunately, the binary format of the time stamp rounds the time value to the nearest 1/128 of a second.  The two periods beat against one another, causing the loss of three time stamps each second.  Additional timing uncertainties due to delays in the detector electronics have not been measured, but are assumed to be on the order of a few milliseconds.  For most time-tag observations, for which time stamps are recorded only once per second, these effects can safely be ignored.

Raw time-tag files are constructed by assigning the value of the most recent time stamp, in units of seconds from the exposure start time, to each subsequent photon event.  The frequency of these time markers determines the temporal resolution of the data.  Photon-arrival times are not modified by the pipeline: values are UTC as assigned by the IDS.  In particular, photon-arrival times {\em are not}\/ converted to a heliocentric scale.

\subsection{Combining Data from Multiple Exposures}\label{sec_multiple}

For each \fuse\/ observation, OPUS combines data from individual exposures into a set of observation-level spectra, as described in \S\ \ref{obs_files}.  While these files are sufficient for many projects, other projects may benefit from specialized data processing.  Here are some points to keep in mind when combining \fuse\/ data from multiple exposures:  
For bright targets, the goal is to maximize spectral resolution, so it is important to align precisely the spectra from individual exposures before combining them.  The wavelength zero points of segments A and B are consistent across each of the \fuse\/ detectors (\S\ \ref{sec_wavecal}), so shifts measured for one detector segment can safely be applied to the other.  For observations made before 2005 July, the LiF1 spectrum is likely to have the most accurate wavelength scale, so it serves as the standard for the other three channels.  For later observations, the LiF2 spectra are likely to be the most accurate.  A procedure to cross-correlate and shift spectra by hand is described in \anchor{http://archive.stsci.edu/fuse/cookbook.html}{{\em The FUSE Data Analysis Cookbook}}.  When cross-correlating the spectra of point-source targets, it is important to exclude regions contaminated by airglow features, as their motions are unlikely to track those of the target.
For faint targets, the goal is to optimize the fidelity of the background model by maximizing the signal-to-noise ratio on background regions of the detector, a goal achieved by combining the IDFs from multiple exposures before extracting the spectra.  
A variety of C- and IDL\footnote{IDL is a registered trademark of ITT Corporation for their Interactive Data Language software.}-based tools to perform these and other data-analysis tasks has been generated by the \fuse\/ project.  Software and documentation are available from MAST.

\acknowledgments

We acknowledge with gratitude the efforts of those who contributed to the design and implementation of initial versions of the CalFUSE pipeline and its associated calibration files: G.~A. Kriss, E.~M. Murphy, J. Murthy, W.~R. Oegerle, and K.~C. Roth.  This research has made use of the Multimission Archive at the Space Telescope Science Institute (MAST).  STScI is operated by the Association of Universities for Research in Astronomy, Inc., under NASA contract NAS5-26555. Support for MAST for non-HST data is provided by the NASA Office of Space Science via grant NAG5-7584 and by other grants and contracts.  This work is supported by NASA contract NAS5-32985.

{\it Facility:} \facility{FUSE}

%\appendix
 \renewcommand{\thesection}{A}
 \renewcommand{\thesubsection}{A-\arabic{subsection}}
  % redefine the command that creates the equation no.
  \setcounter{section}{0}  % reset counter 
  \setcounter{subsection}{0}  % reset counter 
%  \section*{APPENDIX}  % use *-form to suppress numbering
\begin{center}{{\small APPENDIX}}\end{center}
  
\section{File Formats}\label{formats}

All \fuse\/ data are stored as FITS files \citep{Hanisch:01} containing one or more Header + Data Units (HDUs).  The first is called the primary HDU (or HDU~1); it consists of a header and an optional N-dimensional image array.  The primary HDU may be followed by any number of additional HDUs, called ``extensions.''  Each extension has its own header and data unit.  \fuse\/ employs two types of extensions, image extensions (a 2-dimensional array of pixels) and binary table extensions (rows and columns of data in binary representation).  CalFUSE uses the \anchor{http://heasarc.gsfc.nasa.gov/docs/software/fitsio/fitsio.html}{CFITSIO} subroutine library \citep{Pence:99} to read and write FITS files.

%\clearpage

\begin{deluxetable}{lll}
\tablecolumns{3}
\tablewidth{0pt}
\tablecaption{Format of Raw Time-Tag Files\label{ttag}}
\tablehead{
\colhead{Array Name} & \colhead{Format} & \colhead{Description}
}
\startdata
\cutinhead{Primary Header-Data Unit (HDU 1)}
\multicolumn{3}{l}{Header only.  Keywords contain exposure-specific information.} \\
\cutinhead{HDU 2: Photon Event List}
TIME     &       FLOAT    & Photon arrival time (seconds) \\
X    &        SHORT &   Raw X position (0--16383) \\
Y    &        SHORT &   Raw Y position (0--1023) \\
PHA        &     BYTE  &   Pulse height (0--31)\\
\cutinhead{HDU 3: Good-Time Intervals}
START	&	DOUBLE	&	GTI start time (seconds) \\
STOP	&	DOUBLE	&	GTI stop time (seconds) 
\enddata
\tablecomments{Times are relative to the exposure start time, stored in the header keyword EXPSTART.}
\end{deluxetable}

\subsection{Raw Time-Tag and Histogram Files}\label{raw_format}

\fuse\/ raw data files are generated by OPUS using both data downlinked by the telescope and information from the \fuse\/ Mission Planning Database (\S\ \ref{sec_OPUS}).  Information regarding the target, exposure times, instrument configuration, and engineering parameters is stored in a series of header keywords in the primary HDU.  All header keywords are described in the \anchor{http://archive.stsci.edu/fuse/dhbook.html#FITSHeader}{{\em FUSE Instrument and Data Handbook}}.  In raw time-tag files (Table \ref{ttag}), the primary HDU consists of a header only, with no associated image array.  HDU~2 contains the photon-event list, with arrival time (in seconds from the exposure start time), raw detector coordinates, and pulse height for each event in turn.  HDU~3 lists good-time intervals (GTIs) calculated by OPUS.  Raw time-tag file names end with the suffix ``ttagfraw.fit.''  They can be as large as 10--20 MB for the brightest targets.  

The data in raw histogram files (suffix ``histfraw.fit'') are stored as a series of image extensions (Table \ref{hist}).  The primary HDU contains the same header keywords as time-tag files, along with a small ($8 \times 64$ pixel) image called the Spectral Image Allocation (SIA) table. The SIA table is used to map regions of the detector to on-board memory. Each element in the SIA table corresponds to a $2048 \times 16$ pixel region on a detector segment.  If the element is set to 1, the photons from the corresponding region are saved; if 0, they are discarded.  Additional image extensions follow, each containing the binned image of some region of the detector; these regions may overlap.  In general, science data are binned by 8 pixels in Y and unbinned in X; binning factors for each exposure are stored in the header keywords SPECBINX and SPECBINY.  While the format given in Table \ref{hist} is standard, any number of image extensions may be present in a histogram file.  Raw histogram data files are 1--1.5 MB in size.

\subsection{Housekeeping and Jitter Files}\label{sec_jitter}

For each exposure, a single housekeeping file is generated by OPUS from engineering data supplied by the spacecraft  (\S\ \ref{sec_OPUS}).  Housekeeping files (suffix ``hskpf.fit'') contain 62 arrays, including spacecraft pointing information, detector voltage levels, and various counter values, in a single binary table extension.  Arrays are tabulated once per second, though most parameters are updated only once every 16 seconds.  Only a few of the housekeeping arrays are employed by the pipeline.  The detector high voltage and LiF, SiC, FEC, and AIC counter arrays are used to populate the corresponding arrays in the IDF timeline table (\S\ \ref{idf_format}).

From pointing information in the housekeeping file, OPUS derives a jitter file (suffix ``jitrf.fit'') consisting of a single binary table extension with 4 columns: TIME, DX, DY and TRKFLG. The time refers to the elapsed time (in seconds) from the start of the exposure. Since the engineering data commonly begin up to a minute before the exposure, the first few entries of this array are negative. DX and DY are the offsets along the X (dispersion) and Y (cross-dispersion) directions in arc seconds. These offsets are defined relative to the commanded position of the telescope (presumably the target coordinates). Finally, TRKFLG is the tracking quality flag.  Its value is $-1$ if the spacecraft is not tracking properly and 0 if tracking information is unavailable.  Values between 1 and 5 represent increasing levels of fidelity for DX and DY.

Additional details regarding the contents and format of the housekeeping and jitter files are provided in \anchor{http://archive.stsci.edu/fuse/dhbook.html}{{\em The FUSE Instrument and Data Handbook.}}

\subsection{Intermediate Data File (IDF)}\label{idf_format}

The IDF (suffix ``idf.fit'') contains three FITS binary table extensions; their contents are listed in Table \ref{idf}.  The file's primary header-data unit (HDU~1) is copied directly from the raw data file.  (For histogram data, the SIA table is discarded.)  Various keywords are populated by the initialization routine (\S\ \ref{init}) and by subsequent pipeline modules.  The first binary-table extension (HDU~2) contains the photon events themselves.  For time-tag data, the TIME, XRAW, YRAW, and PHA arrays are copied from the raw data file, and the WEIGHT array is initialized to 1.0.  For histogram data, each image pixel is mapped back to its coordinates on the full detector, which are recorded in the XRAW and YRAW arrays.   The WEIGHT array is initialized to the number of photon events in the pixel.  Zero-valued pixels are ignored.  Histogram data are not ``unbinned.''  Each entry of the TIME array is set to the midpoint of the exposure and each entry of the PHA array to 20.  (Both arrays are subsequently modified.)

The first pipeline module (\S\ \ref{convert_to_farf}) corrects for various detector effects; it scales the WEIGHT array to correct for detector dead time and populates the XFARF and YFARF arrays.  (The flight alignment reference frame represents the output of an ideal detector.)  Each photon event is assigned to one of six aperture-channel combinations or to the background (Table \ref{channel}) and a corresponding code is written to the CHANNEL array (\S\ \ref{remove_motions}).  After corrections for mirror, grating, and spacecraft motions, the photon's final coordinates are recorded in the X and Y arrays.  Though floating-point arrays, XFARF, YFARF, X, and Y are written to the IDF as arrays of 8-bit integers using the FITS TZERO and TSCALE keywords.  This process effectively rounds each element of XFARF and X to the nearest 0.25 of a detector pixel and each element of YFARF and Y to the nearest 0.1 of a detector pixel.  

The screening routines (\S\ \ref{sec_screen}) use information from the timeline table (described below) to identify photons that violate pulse-height limits, limb-angle constraints, \etc\  ``Bad'' photons are not deleted from the IDF, but merely flagged.  Flags are stored as single bits in an 8-bit byte.  We use two sets of flags, TIMEFLGS for time-dependent and LOC\_FLGS for location-dependent effects (Table \ref{flags}).  For each bit, a value of 0 indicates that the photon is ``good,'' except for the day/night flag, for which 0 = night and 1 = day.  It is possible to modify these flags without re-running the pipeline.  For example, one could exclude day-time photons or include data taken close to the earth limb.

The LAMBDA array contains the heliocentric wavelength assigned to each photon (\S\ \ref{wavelength}), and the ERGCM2 array records its ``energy density'' in units of erg cm$^{-2}$ (\S\ \ref{flux}).  To convert an extracted spectrum to units of flux, one must divide by the exposure time and the width of an output spectral bin.

The second extension (HDU~3) is a list of good-time intervals (GTIs).  The initial values are copied from the raw data file, but they are modified by the pipeline once the various screening routines have been run.  By convention, the START value of each GTI corresponds to the arrival time of the first photon in that interval.  The STOP value is one second later than the arrival time of the last photon in that interval.  The length of the GTI is thus STOP$ - $START.

The third extension (HDU~4) is called the timeline table.  It contains status flags and spacecraft and detector parameters used by the pipeline.  An entry in the timeline table is created for each second of the exposure.  For time-tag data, the first entry corresponds to the time of the first photon event, and the final entry to the time of the final photon event plus one second.  (Should an exposure's photon-arrival times purport to exceed 55 ks, we create timeline entries only for each second in the good-time intervals.)  For histogram data, the first element of the TIME array is set to zero and the final element to EXPTIME+1 (where EXPTIME is the exposure duration computed by OPUS).  Because we require that EXPTIME equal both $\Sigma$ (STOP$ - $START), summed over all entries in the GTI table, and the number of good times in the timeline table, we must flag the final second of each GTI as bad.  No photons are associated with the STOP time of a GTI.

Only the day/night and OPUS flags of the STATUS\_FLAGS array are populated when the IDF is created; the other flags are set by the various screening routines (\S\ \ref{sec_screen}).  The elements of the TIME\_SUNSET, TIME\_SUNRISE, LIMB\_ANGLE, LONGITUDE, LATITUDE, and ORBITAL\_VEL arrays are computed from the orbital elements in the FUSE.TLE file.  The HIGH\_VOLTAGE array is populated with values from the housekeeping file.
The LIF\_CNT\_RATE and SIC\_CNT\_RATE arrays are initially populated with values derived from the LiF and SiC counter arrays in the housekeeping file.  For time-tag data, these arrays are eventually updated with the actual count rates within the target aperture, excluding regions contaminated by airglow.  The FEC\_CNT\_RATE and AIC\_CNT\_RATE, described in \S\ \ref{deadtime}, are also derived from counter arrays in the housekeeping file.  For time-tag data, the BKGD\_CNT\_RATE array is populated by the burst-rejection routine (\S\ \ref{screen_burst}) and represents the count rate in pre-defined background regions of the detector, excluding airglow features.  The array is not populated for histogram data.  The YCENT\_LIF and YCENT\_SIC arrays trace the centroid of the target spectra with time {\it before} motion corrections are applied.  These two arrays are not used by the pipeline.

Raw time-tag files (\S\ \ref{raw_format}) employ the standard FITS binary table format, listing TIME, X, Y, PHA for each photon event in turn. The intermediate data files have a slightly different format, listing all of the photon arrival times, then the X coordinates, then the Y coordinates. Formally, the table has only one row, and each element of the table is an array.  (To use the STSDAS terminology, IDFs are written as 3-D tables.)  The MDRFITS function from the \anchor{http://idlastro.gsfc.nasa.gov/}{IDL Astronomy User's Library} \citep{Landsman:93} can read both file formats; some older FITS readers cannot. Note that, because HDUs 2 and 4 of the IDFs contain floating-point arrays stored as shorts (using the TZERO and TSCALE keywords), calls to MRDFITS must include the keyword parameter FSCALE.

\subsection{Bad-Pixel Maps (BPM Files)}\label{sec_bpm_format}

The BPM files (suffix ``bpm.fit''; \S\ \ref{sec_bpm}) consist of a single binary table extension.  Its format is similar to that of the IDF, but it contains only five columns: X, Y, CHANNEL, WEIGHT, and LAMBDA.  The WEIGHT column, whose values range from 0 to 1, represents the fraction of the exposure that each pixel was affected by a dead spot.  The BPM files are not archived, but can be generated from the IDF and jitter file using pipeline software available from MAST.

\subsection{Extracted Spectral Files}\label{spec_format}

Extracted spectra (suffix ``fcal.fit''; \S\ \ref{extract_spectra}) are stored in a single binary table extension.  Its contents are presented in Table \ref{spec}.  Note that the spectra are binned in wavelength.  The bin size can be set by the user, but the default is 0.013 \AA, which corresponds to about 2 detector pixels or about one-fourth of a spectral resolution element.  The WAVE array records the central wavelength of each spectral bin.  For time-tag data, the COUNTS array represents the total of all (raw) photon events assigned to the target aperture.  For histogram data, the COUNTS array is simply the WEIGHTS array divided by the mean dead-time correction for the exposure.  If optimal extraction is performed, the values of the FLUX, ERROR, WEIGHTS, and BKGD arrays are determined by that algorithm.  As a result, the ratio of WEIGHTS to COUNTS is constant only for histogram data.  The QUALITY array records the percentage of the extraction window containing valid data.  It is 100 if no bad pixels fell within the wavelength bin, 0 if the entire bin was lost to bad pixels.

%%%%%%%%%%%%% REFERENCES  %%%%%%%%%%%%%%%

%\clearpage 

%%%%%%%%%%%%%%% FIGURES  %%%%%%%%%%%%%%%%

\clearpage

%\clearpage

%\clearpage

%\clearpage

%\clearpage

%\clearpage

%\clearpage

%\clearpage

%\clearpage

%\clearpage

%\clearpage

%\clearpage

%\clearpage

%%%%%%%%%%%%%%% TABLES %%%%%%%%%%%%%%%%%

\clearpage

\begin{deluxetable}{clc}
\tablecolumns{3}
\tablewidth{240pt}
\tablecaption{Format of Raw Histogram Files\label{hist}}
\tablehead{
&&\colhead{Image Size\tablenotemark{a}}\\
\colhead{HDU} & \colhead{Contents} & \colhead{(binned pixels)}
}
\startdata
1\tablenotemark{b} & SIA Table\tablenotemark{c} & $8 \times 64$\\ 
2 & SiC Spectral Image & (12--20) $\times$ 16384\\
3 & LiF Spectral Image & (12--20) $\times$ 16384\\
4 & Left Stim Pulse & 2 $\times$ 2048\\
5 & Right Stim Pulse & 2 $\times$ 2048
\enddata
\tablenotetext{a}{Quoted image sizes assume the standard histogram binning: by 8 pixels in Y, unbinned in X.  Actual binning factors are given in the primary file header.}
\tablenotetext{b}{Header keywords of HDU 1 contain exposure-specific information.}
\tablenotetext{c}{The SIA table describes which regions of the detector are included in the file.}
\tablecomments{While this table describes the format of a typical raw histogram file, any number of HDUs are allowed.}
\end{deluxetable}

%\clearpage

\begin{deluxetable}{lll}
\tablecolumns{3}
\tablewidth{0pt}
\tablecaption{Format of Intermediate Data Files\label{idf}}
\tablehead{
\colhead{Array Name} & \colhead{Format} & \colhead{Description}
}
\startdata
\cutinhead{Primary Header-Data Unit (HDU 1)}
\multicolumn{3}{l}{Header only.  Keywords contain exposure-specific information.} \\
\cutinhead{HDU 2: Photon Event List}
TIME     &       FLOAT    & Photon arrival time (seconds)\\
XRAW    &        SHORT &   Raw X coordinate (0--16383) \\
YRAW    &        SHORT &   Raw Y coordinate (0--1023) \\
PHA        &     BYTE  &   Pulse height (0--31) \\
WEIGHT   &       FLOAT &   Photons per binned pixel for HIST data, \\
& & initially 1.0 for TTAG data \\
XFARF    &       FLOAT &   X coordinate in geometrically-corrected frame \\
YFARF    &       FLOAT  &  Y coordinate in geometrically-corrected frame \\
X          &     FLOAT  &  X coordinate after motion corrections \\
Y        &       FLOAT  &  Y coordinate after motion corrections \\
CHANNEL     &    BYTE  &   Aperture+channel ID for the photon (Table \ref{channel}) \\
TIMEFLGS    &    BYTE  &   Time flags (Table \ref{flags}) \\
LOC\_FLGS    &    BYTE  &   Location flags (Table \ref{flags}) \\
LAMBDA       &   FLOAT  &  Wavelength of photon (\AA)  \\
ERGCM2   &      FLOAT  &  Energy density of photon (erg cm$^{-2}$) \\
\cutinhead{HDU 3: Good-Time Intervals}
START	&	DOUBLE	&	GTI start time (seconds)  \\
STOP	&	DOUBLE	&	GTI stop time (seconds) \\
\cutinhead{HDU 4: Timeline Table}
TIME     &       FLOAT     &      Seconds from exposure start time \\
STATUS\_FLAGS &   BYTE     &       Status flags \\
TIME\_SUNRISE  &  SHORT   &        Seconds since sunrise \\
TIME\_SUNSET &    SHORT   &        Seconds since sunset \\
LIMB\_ANGLE   &   FLOAT      &    Limb angle (degrees)  \\
LONGITUDE    &   FLOAT     &     Spacecraft longitude (degrees)  \\
LATITUDE     &   FLOAT       &   Spacecraft latitude (degrees)  \\
ORBITAL\_VEL  &  FLOAT     &     Component of spacecraft velocity \\
\multicolumn{2}{}{} & in direction of target (km/s)\\
HIGH\_VOLTAGE &   SHORT   &        Detector high voltage (unitless) \\
LIF\_CNT\_RATE &   SHORT     &      LiF count rate (counts/s) \\
SIC\_CNT\_RATE &   SHORT    &       SiC count rate (counts/s) \\
FEC\_CNT\_RATE &   FLOAT    &       FEC count rate (counts/s) \\
AIC\_CNT\_RATE  &  FLOAT      &     AIC count rate (counts/s)  \\
BKGD\_CNT\_RATE &  SHORT    &       Background count rate (counts/s) \\
YCENT\_LIF    &   FLOAT     &      Y centroid of LiF target spectrum (pixels)\\
YCENT\_SIC   &    FLOAT    &       Y centroid of SiC target spectrum (pixels)
\enddata
\tablecomments{Times are relative to the exposure start time, stored in the header keyword EXPSTART.  To conserve memory, floating-point values are stored as shorts (using the FITS TZERO and TSCALE keywords) except for TIME, WEIGHT, LAMBDA and ERGCM2, which remain floats.}
\end{deluxetable}

%\clearpage

\begin{deluxetable}{lcc}
\tablecolumns{3}
\tablewidth{0pc} 
\tablecaption{Aperture Codes for IDF CHANNEL Array\label{channel}}
\tablehead{   
\colhead{Aperture} & \colhead{LiF}   & \colhead{SiC}}
\startdata 
HIRS     &   1    &    5   \\
MDRS    &    2    &   6   \\
LWRS    &    3    &    7   \\
\multicolumn{2}{l}{Not in an aperture}	& 0
\enddata
\end{deluxetable}

%\clearpage

\begin{deluxetable}{ll}
\tablecolumns{2}
\tablewidth{0pc} 
\tablecaption{Bit Codes for IDF Time and Location Flags\label{flags}}
\tablehead{   
\colhead{Bit} & \colhead{Value}}
\startdata 
\multicolumn{2}{c}{Time Flags} \\
\cline{1-2} \\
        8 & User-defined bad-time interval \\
        7 & Jitter (target out of aperture) \\
        6 & Not in an OPUS-defined GTI {\it or} \\
           &  Photon arrival time unknown \\
        5 & Burst  \\
        4 & High voltage reduced\\
        3 & SAA \\
        2 & Limb angle \\
        1 & Day/Night flag (N = 0, D = 1) \\
\cutinhead{Location Flags}
	8 & Not used  \\
	7 & Fill data (histogram mode only) \\
	6 & Photon in bad-pixel region  \\
        5 & Photon pulse height out of range \\
        4 & Right stim pulse \\
        3 & Left stim pulse \\
        2 & Airglow feature \\
        1 & Not in detector active area
\enddata
\tablecomments{Flags are listed in order from most- to least-significant bit.}
\end{deluxetable}

%\clearpage

\begin{deluxetable}{lll}
\tablecolumns{3}
\tablewidth{0pt}
\tablecaption{Format of Extracted Spectral Files\label{spec}}
\tablehead{
\colhead{Array Name} & \colhead{Format} & \colhead{Description}
}
\startdata
\cutinhead{Primary Header-Data Unit (HDU 1)}
\multicolumn{3}{l}{Header only.  Keywords contain exposure-specific information.} \\
\cutinhead{HDU 2: Extracted Spectrum}
WAVE     &       FLOAT    & Wavelength (\AA)\\
FLUX     &       FLOAT    & Flux (\flux)\\
ERROR     &       FLOAT    & Gaussian error (\flux)\\
COUNTS     &       INT    & Raw counts in extraction window \\
WEIGHTS     &       FLOAT    & Raw counts corrected for dead time \\
BKGD     &       FLOAT    & Estimated background in extraction window (counts)\\
QUALITY     &       SHORT    & Percentage of window used for extraction (0--100)
\enddata
\end{deluxetable}

\end{document}